\documentclass{aa} 
\usepackage{graphicx}
\usepackage{lscape}
\usepackage{longtable}
\usepackage{arydshln}
\usepackage{amsmath}
\usepackage{txfonts}
\usepackage{natbib}
\usepackage[htt]{hyphenat}
\usepackage{breqn}
\usepackage[normalem]{ulem}

\bibpunct{(}{)}{;}{a}{}{,} 

\begin{document}

\title{AB Aur, a Rosetta stone for studies of planet formation (IV): C/O estimates from CS and SO interferometric observations.} 

\author{P. Rivi\`ere-Marichalar \inst{1}
\and A. Fuente \inst{2}
\and R. le Gal \inst{3,4}
\and R. Neri  \inst{3}
\and G. Esplugues \inst{1}
\and D. Semenov \inst{5,6}
\and R. Teague \inst{7}
\and Alejandro Santamar\'ia-Miranda \inst{1}
}

\institute{Observatorio Astron\'omico Nacional (OAN,IGN), Calle Alfonso XII, 3. 28014 Madrid, Spain 
                   \email{p.riviere@oan.es}
        \and  Centro de Astrobiolog\'ia (CAB), INTA-CSIC, Carretera de Ajalvir Km. 4, Torrej\'on de Ardoz, 28850 Madrid, Spain 
        \and Institut de Radioastronomie Millim\'etrique, 300 rue de la Piscine, F-38406 Saint-Martin d'H\`eres, France 
	\and Institut de Plan\'etologie et d'Astrophysique de Grenoble (IPAG), Universit\'e Grenoble Alpes, CNRS, 38000 Grenoble, France
        \and Max-Planck-Institut f\"{u}r Astronomie (MPIA), K\"{o}nigstuhl 17, D-69117 Heidelberg, Germany 
	\and Department of Chemistry, Ludwig-Maximilians-Universit\"{a}t, Butenandtstr. 5-13, D-81377 M\"{u}nchen, Germany
	\and Department of Earth, Atmospheric, and Planetary Sciences, Massachusetts Institute of Technology, Cambridge, MA 02139, USA  
}

\authorrunning{Rivi\`ere-Marichalar et al.}
\titlerunning{AB Aur, a Rosetta stone for studies of planet formation (IV)}
\date{}

 \abstract 
{Protoplanetary disks are the birthplace of planets. As such, they set the initial chemical abundances available for planetary atmosphere formation. Thus, studying elemental abundances, molecular compositions, and abundance ratios in protoplanetary disks is key to linking planetary atmospheres to their formation sites. }
{We aim to derive the sulfur abundance and the C/O ratio in the AB Aur disk using interferometric observations of CS and SO.}
{New NOEMA observations of CS 3-2 towards AB Aur are presented. We used velocity-integrated intensity maps to determine the inclination and position angles. Keplerian masks were constructed for all observed species to assess the presence of non-Keplerian motions. We use the CS/SO ratio to study the C/O ratio. We compare our present and previous interferometric observations of AB Aur with a NAUTILUS disk model to gain insight into the S elemental abundance and C/O ratio. }
{We derive an observational CS/SO ratio ranging from 1.8 to 2.6. Only NAUTILUS models with C/O$\gtrsim$1 can reproduce such ratios. The comparison with models points to strong sulfur depletion, with [S/H]=8$\times$10$\rm ^{-8}$, but we note that no single model can simultaneously fit all observed species. }
{}

\keywords{Astrochemistry -- ISM: abundances  -- ISM: molecules --
   stars: formation}

\maketitle
 
\begin{figure*}[t!]
\begin{center}
  \includegraphics[width=1\textwidth, trim=0mm 10mm 0mm 0mm, clip]{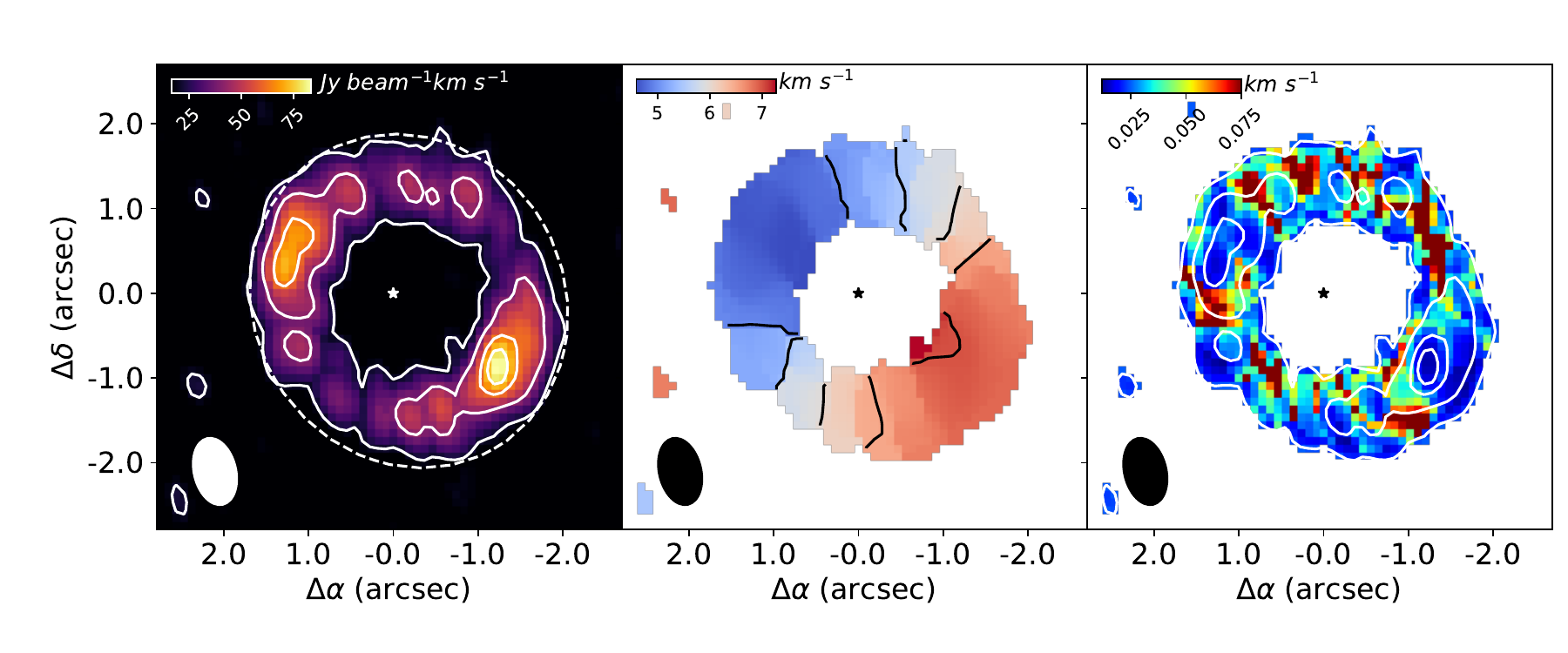}
  \caption{From left to right: zeroth-, first, and second-moment map of CS 3-2 observed with NOEMA. The first- and second-moment maps were computed using a 3$\sigma$ threshold to avoid noisy channels. Contours in the zeroth- and second-moment maps are shown for 0.25, 0.5, 0.75, and 0.9 of the integrated intensity map peak. The white dashed ellipse in the zeroth-moment map shows the area used to compute the integrated spectrum shown in Fig. \ref{Fig:CS_spectrum}.}
 \label{Fig:CS_moments}
\end{center}
\end{figure*}

\section{Introduction} 
Planets form within protoplanetary disks and inherit much of their chemical composition from them \citep{Oberg2021}. Thus, characterizing both the physical properties and chemical content of protoplanetary disks is key to understanding the formation and composition of planetary atmospheres. However, our understanding of the chemistry in these systems remains incomplete. Furthermore, while planets are assembled in the disk's mid-plane, most molecular species trace the disk atmosphere's upper layers or the so-called warm molecular layer \citep{Aikawa1999}. Thus, finding species that emit close to the mid-plane is essential, as it will allow us to study the physics and chemistry around the planet-forming region. The composition of planetary atmospheres traces the elemental composition in the planet-formation site and can be used to locate the formation site and understand its chemical history \citep{Oberg2011}. Among all the elements, carbon, hydrogen, oxygen, nitrogen, phosphorus, and sulfur (the so-called CHONPS)  are particularly important due to their role in biochemistry. Sulfur is the tenth most abundant element in the universe. Its abundance toward the Sun is S/H$\sim$1.5$\times$10$^{-5}$ \citep[]{Asplund2009} which is consistent with the value found in Orion B stars \citep[S/H$\sim$1.4$\times$10$^{-5}$,][]{Daflon2009}. However, sulfur chemistry in the ISM is poorly understood, and sulfur-bearing molecules are not as abundant as expected. A value that is close to the cosmic one is observed in photon-dominated regions (PDRs)  \citep{Goicoechea2021b, Fuente2024}, while strong depletion is observed in dense molecular gas \citep{Vastel2018, Fuente2023}. Star-forming regions show evidence of a trend in sulfur depletion with age, with more evolved systems showing larger sulfur depletion \citep{Hily-Blant2022, Esplugues2022}. A natural explanation is that sulfur is locked in ice on the surfaces of dust grains. However, only OCS has been confirmed thus far in the ice phase \citep{Palumbo1995, McClure2023}, while \cite{McClure2023} reported hints of SO$\rm _2$ ice emission. Hence, to understand sulfur chemistry, we must compare observations of gas-phase species with models. 

Protoplanetary disks are expected to reprocess the molecular content inherited from their natal cloud. They offer a unique laboratory to understand the huge differences in sulfur abundance between the different phases of the ISM.  However, the study of the sulfur abundance in protoplanetary disks has been restricted to a few systems and has provided few detections. The only species widely detected is CS \citep{Dutrey2011, Guilloteau2012, Pacheco2015, Guilloteau2016, Teague2018b, Podio2020, Podio2020b, Rosotti2021, Riviere2021, Nomura2021}.  H$_2$CS was detected in MWC 480 \citep{LeGal2019}, and H$_2$S was detected in GG Tau  \citep{Phuong2018}, AB Aur \citep{Riviere2022}, and four protoplanetary disks in Taurus \citep{Riviere2021}. SO has been detected in a few PPDs \citep{Pacheco2016,Booth2018,Booth2021,LeGal2021,Huang2024}, and SO$\rm _2$ has been observed toward IRS 48 \citep{Booth2021}. \cite{Phuong2021} detected CCS towards GG Tau using NOEMA, but they could not track the spatial distribution due to the low S/N. The authors could not reproduce the observed column density of CCS and other sulfur-bearing species simultaneously. This suggests that current sulfur astrochemical networks may be incomplete \citep{Navarro2020}.

A key parameter for the chemistry of protoplanetary systems is the C/O ratio, which impacts the composition of planetary atmospheres. As a result, the C/O ratio of planetary atmospheres can be used to constrain the distance from the central protostar at which planets were formed \citep{Oberg2011}. Models predict that the C/O ratio will experience enhancements with radius as we cross the snowlines of different species \citep{Oberg2011}.One way to derive the C/O ratio is by comparing the observed CS/SO ratio with astrochemical model predictions. However, this method is limited by the small number of SO detections in protoplanetary disks. The few SO detections \citep{Pacheco2016,Booth2018,Booth2021,Huang2023,Law2023} so far suggest that there is a population of oxygen-rich disks \citep{Huang2024}. There is a prevalence of transitional Herbig Ae and Be (HAeBe) stars among the sources with an SO detection (four out of five), which suggests that the strong irradiation from the host protostar is heating dust grains to the point where SO is desorbed \citep{Booth2021}. 

One of the few systems where SO has been detected so far is AB Aur (RA$\rm _{J2000}$=04:55:45.8459, DEC$\rm _{J2000}$=+30:33:04.291), a well-known Herbig Ae star \citep[spectral type A0-A1][]{Hernandez2004}  that hosts a transitional disk and an embedded planet candidate \citep{Currie2022}. In its last data release, the GAIA catalog provides a distance of 155.9$\rm \pm$0.9 pc to AB Aur \citep{GAIA2020}. We have performed an observational study of molecular emission in AB Aur including HCN, HCO$\rm ^+$, $\rm ^{12}$CO, $\rm ^{13}$CO, C$\rm ^{18}$O, SO, H$\rm _2$CO, and H$\rm _2$S, with radii of emission peaks ranging from  0.9$\arcsec$ to 1.4$\arcsec$ \citep{Riviere2019, Riviere2020, Riviere2022}. In the present paper, we present NOEMA observations of CS 3-2 and use them to study the sulfur content and the C/O ratio of the protoplanetary disk. The paper is organized as follows. In Sec. \ref{Sect:obs_data_red} we describe the data reduction methods. In Sec. \ref{Sect:results}, we present our main results, including accurate estimates of the disk inclination and position angles, radial and azimuthal profiles, and forecasts of the CS column density. In Sec. \ref{Sect:discussion}, we briefly discuss our results regarding the sulfur content and C/O ratio of protoplanetary disks. Finally, we summarize our results in Sec. \ref{Sect:summary}.

\section{Observations and data reduction}\label{Sect:obs_data_red}
\subsection{NOEMA observations}
The observations were conducted using NOEMA under project W21BA (PI: P. Rivi\`ere-Marichalar) in the AB set of configurations, utilizing all twelve antennas between February 20 and March 5, 2022. Observing conditions were generally excellent, with stable and favorable weather throughout the campaign. The atmospheric phase stability ranged between 10$^\circ$ and 40$^\circ$ RMS, and precipitable water vapor values were between 0.5 and 5\,mm.

The NOEMA antennas, equipped with 2SB receivers, were tuned to cover the frequency ranges from 127.737 to 135.781 GHz (LSB) and 143.225 to 151.269 GHz (USB) in both polarizations. The correlator covered the full frequency range with 2 MHz spectral channels and was configured to include an additional 64 MHz high-resolution window targeting the CS 3-2 line. Amplitude and phase calibration were performed using J0438+300 and J0418+380 as calibrators, with flux densities bootstrapped to standard references LkH$\alpha$101 and MWC\,349. Calibration, data reduction, and analysis were conducted using the GILDAS software package. The estimated accuracy of the absolute flux calibration is 15\%.

The continuum image was created by averaging the line-free channels from the LSB and USB. This resulted in a phase self-calibrated image with a thermal noise level of 13.6$\,\mu$Jy beam$^{-1}$ (1$\sigma$). The synthesized beam for the continuum image has a size of $0\farcs79 \times 0\farcs48$ at a position angle of 13$^\circ$, providing adequate resolution to probe the small-scale structures in the AB Aur disk. The derived complex gains from the continuum self-calibration were subsequently applied to the spectral line visibilities, after which the velocity-integrated intensity maps for CS were generated.

\subsection{ALMA observations}
We used ALMA observations from project 2021.1.00690 (PI Dong) to study SO emission. The reduction of these datasets was described in \citep{Dutrey2024}, and we refer the reader to this paper for a description of the data reduction process. 

\begin{figure}[t!]
\begin{center}
  \includegraphics[width=0.5\textwidth]{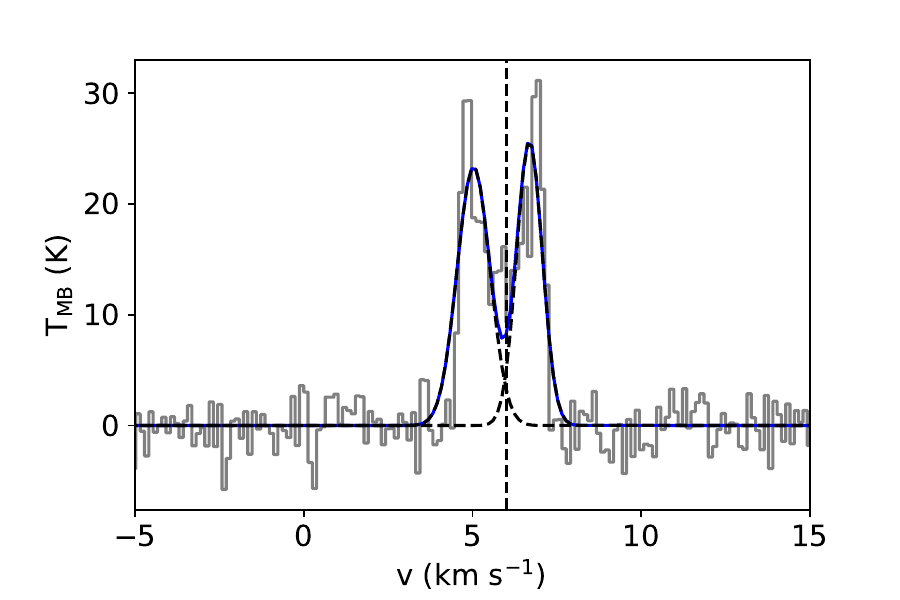}
  \caption{CS spectrum extracted from a 2$\arcsec \times $1.8$\arcsec$ ellipse with PA=237 $^\circ$ centered on the CS emission disk (see Fig. \ref{Fig:CS_moments}, left panel). The blue curve shows a 2-Gaussian fit to the data, with the black dashed curves representing the individual Gaussians.}
 \label{Fig:CS_spectrum}
\end{center}
\end{figure}

\begin{figure}[t!]
\begin{center}
  \includegraphics[width=0.5\textwidth]{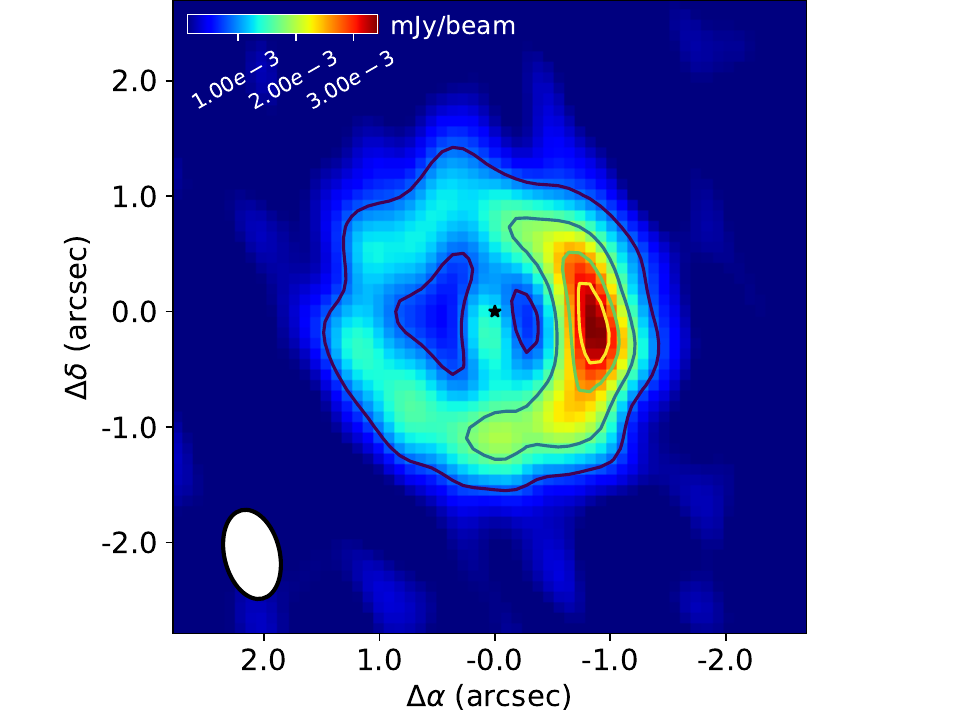}
  \caption{Continuum map at 2 mm. Contours are shown for 0.25, 0.5, 0.75, and 0.9 of the map peak.}
 \label{Fig:cont}
\end{center}
\end{figure}

\section{Results}\label{Sect:results}
In the top left panel of Fig. \ref{Fig:CS_moments}, we show the velocity integrated intensity map (moment zero) of our NOEMA CS 3-2 observations. We also show the corresponding first- (intensity-weighted average velocity) and second-moment (intensity-weighted average velocity dispersion) maps in the center and right panels of Fig. \ref{Fig:CS_moments}. The maps were integrated between 4.3 and 7.3 km s$^{-1}$. The first- and second-moment maps were computed using a 3$\sigma$ threshold in the channels. The velocity-integrated intensity map depicts an elliptical ring of emission (aspect ratio 1.2). The disk extends from 0.9$\arcsec$ to 2$\arcsec$ and has an azimuthal contrast ratio of 2.4. The integrated intensity map shows two local maxima, one close to the dust trap \citep[western peak in continuum image,][]{Fuente2017} and a second at roughly 180$\rm ^\circ$ in azimuth. These two local maxima coincide with regions of low velocity dispersion (see Fig. \ref{Fig:CS_moments}, right panel). The first-moment map (intensity-weighted velocity) exhibits the characteristic velocity pattern of a Keplerian disk. From the velocity-integrated intensity map, we compute a CS 3-2 line flux of (0.35$\rm \pm$0.09) Jy km $\rm s^{-1}$.

Fig. \ref{Fig:CS_spectrum} shows the integrated CS 3-2 spectrum with the characteristic double-peak profile emerging from a disk in Keplerian rotation. Fig. \ref{Fig:CS_spectrum} also includes a Gaussian fit to the observations (blue lines) consisting of two individual Gaussians for the blueshifted and redshifted peaks, with peaks at velocities $\rm v=5.0 ~km~s^{-1}$ and $\rm v=6.7 ~km~s^{-1}$, and line widths (FWHM) 1.15 and 0.89 km s$\rm ^{-1}$, respectively. The center of the spectral profile is located at $\rm \sim$5.86 km/s.

In Fig.  \ref{Fig:cont} we show a continuum emission map at 2 mm built by masking line emission in our spectroscopic data cube. Two components are visible in the map: a compact emission peak at the stellar position and a ring corresponding to the protoplanetary disk. Continuum emission in the ring is observed from 0.6$\arcsec$ out to 1.5$\arcsec$, with an azimuthal contrast ratio of 2.3. The aspect ratio between the major and minor axes of the continuum ring at 2 mm is 1.1.  We computed a flux of (26$\rm \pm$4) mJy at 2 mm.

\begin{figure}[t!]
\begin{center}
  \includegraphics[width=0.5\textwidth]{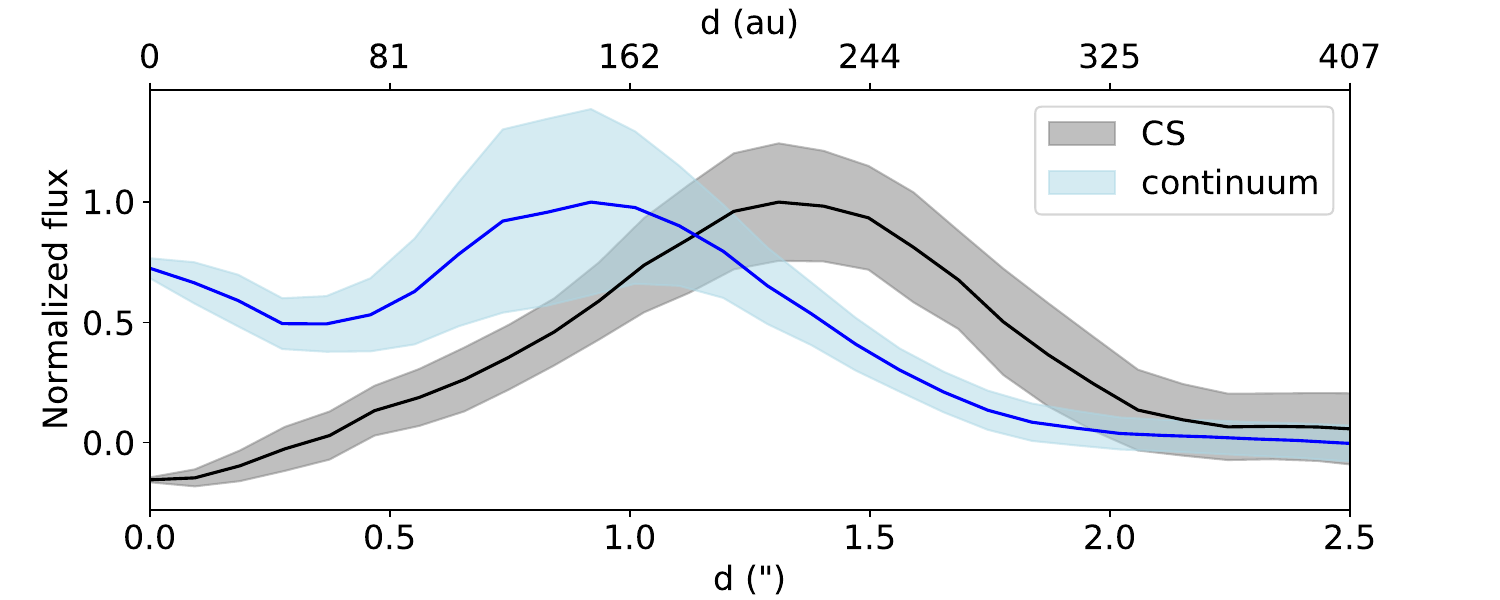}
  \caption{Azimuthally averaged radial profile of the CS 3-2 velocity integrated intensity map (black line) and the continuum at 2 mm (blue line). The shaded areas account for the uncertainties in the profiles.}
 \label{Fig:CS_radprof}
\end{center}
\end{figure}

\subsection{Disk inclination and position angle}\label{Subsect:i_and_PA}
Utilizing the extensive dataset of molecular species observations for this source, we calculated the mean disk inclination and position angles from all available observations. We computed first moment-maps (intensity-weighted velocity) of our CS, HCO$^+$, HCN, $^{13}$CO, C$^{18}$O, H$_2$CO, H$_2$S \citep{Riviere2019, Riviere2020, Riviere2022} interferometric observations using \texttt{bettermoments} \citep{Teague2018, Teague2019} with a 5$\sigma$ clipping. The resulting maps were used to fit the velocity map of each transition using \texttt{eddy} \citep{eddy}, which performs an MCMC exploration of the parameter space. We assumed a distance $d=155.9$ pc \citep{GAIA2020}, a stellar mass $M_*$=2.4 $\rm M_{\odot}$ \citep{vandenAncker1998}, and a system velocity $v_{lsr}$=5.8 km~s$\rm ^{-1}$. Thus, the free parameters in the fit were the coordinates of the phase center ($x_0$ and $y_0$), the inclination angle of the disk ($i$), and its position angle ($PA$). The starting positions were $x_0$=$y_0$=0\arcsec, $i$=24.9$^\circ$, $PA$=217$^\circ$. The results of this exercise are summarized in Table \ref{Tab:eddy_results}. The estimates from the different molecular species are consistent. The mean values and standard deviations are $i=$22.0$^\circ$$\rm \pm$0.5$^\circ$, PA=237.0$^\circ$$\rm \pm$0.7$^\circ$, in good agreement with literature values \citep[see, e.g. ][]{Tang2012}. 

 \begin{table}[]
\caption{\texttt{eddy} fit results.}
\label{Tab:eddy_results}
\centering
\begin{tabular}{llll}
\hline \hline
Species & Transition & i ($\rm ^{\circ}$) & PA ($\rm ^{\circ}$) \\
\hline
C$^{18}$O & 2-1 & 21.80$\rm \pm$0.01 & 236.74$\rm \pm$0.02 \\
$^{13}$CO & 2-1 & 21.95$\rm \pm$0.01 & 237.77$\rm \pm$0.03 \\
H$_2$CO & 3$\rm  _{03}$-2$\rm _{02}$& 22.6$\rm \pm$0.01 & 237.87$\rm \pm$0.04 \\
H$_2$S & 1$\rm  _{10}$-1$\rm _{01}$ & 22.83$\rm \pm$0.05 &  235.5$\rm \pm$0.2\\
HCO$^+$ & 3-2 & 21.68$\rm \pm$0.02 & 236.84$\rm \pm$0.04  \\
HCN & 3-2 & 21.38$\rm \pm$0.2 &  237.21$\rm \pm$0.05\\
CS & 3-2 & 22.01$\rm \pm$0.02 & 236.9$\rm \pm$0.05\\
\hline
\end{tabular}
\end{table}

\subsection{Radial profile}
In Fig. \ref{Fig:CS_radprof}, we show the azimuthally averaged radial profile of our CS 3-2 velocity-integrated intensity map after deprojection assuming i=22.1$^\circ$ and PA=240$^\circ$, as obtained in Sect. \ref{Subsect:i_and_PA}. The profile peaks at $\rm \sim$1.3$\rm \arcsec$, or $\rm \sim$212 au at the distance to AB Aur \citep[155.9 pc,][]{GAIA2020}. In comparison, SO peaks at 1.4$\arcsec$ \citep[$\rm \sim$230 au][]{Riviere2020}, while $\rm H_2S$ peaks at 1.2$\arcsec$ \citep[$\rm \sim$195 au][]{Riviere2022}. Fig. \ref{Fig:CS_radprof} also includes the radial profile of continuum emission at 2mm (blue line and shaded area), which peaks at 0.96$\arcsec$ (156 au) from the center, a difference of almost 60 au concerning CS 3-2 emission. We note that the profile is not Gaussian, but a Gaussian fit provides a good estimate of the CS ring width through the FWHM of the  Gaussian fit. The best fit has a FWHM of (0.86$\rm \pm$0.02)$\arcsec$, or $\rm \sim (141 \pm$3) au.

\begin{figure*}[t!]
\begin{center}
  \includegraphics[width=0.3\textwidth, trim=0mm 0mm 0mm 0mm, clip]{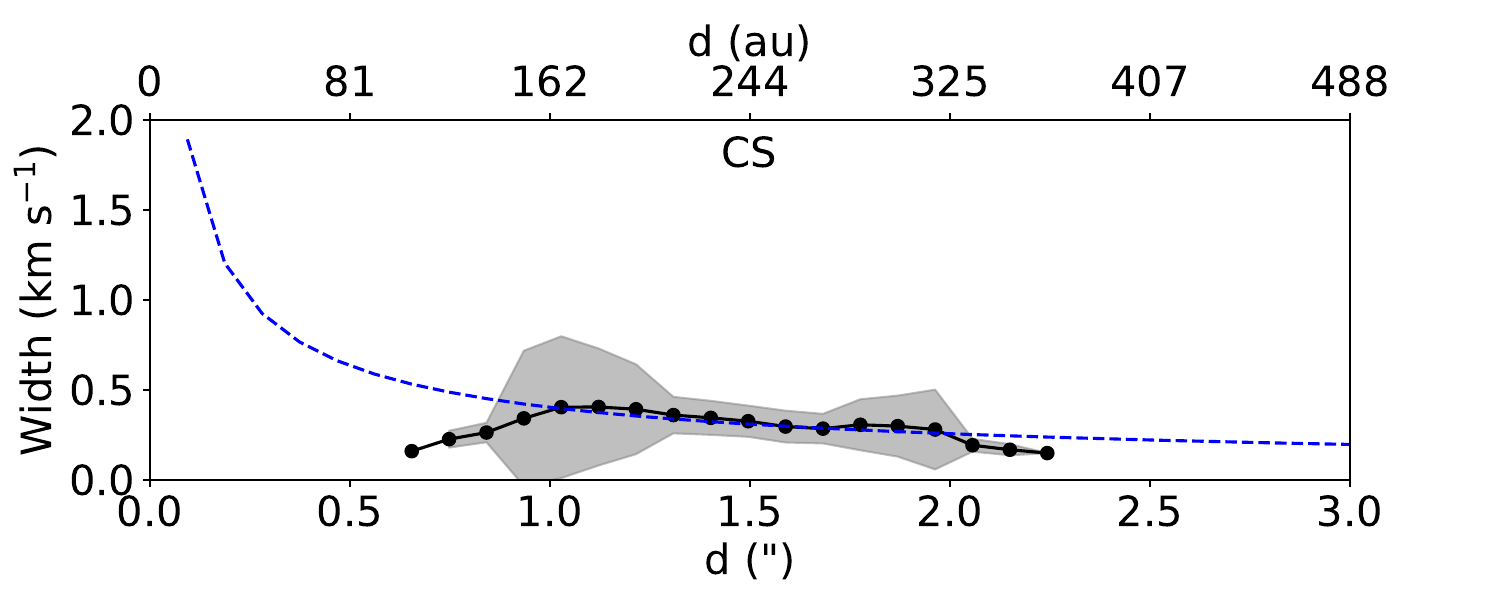}
  \includegraphics[width=0.3\textwidth, trim=0mm 0mm 0mm 0mm, clip]{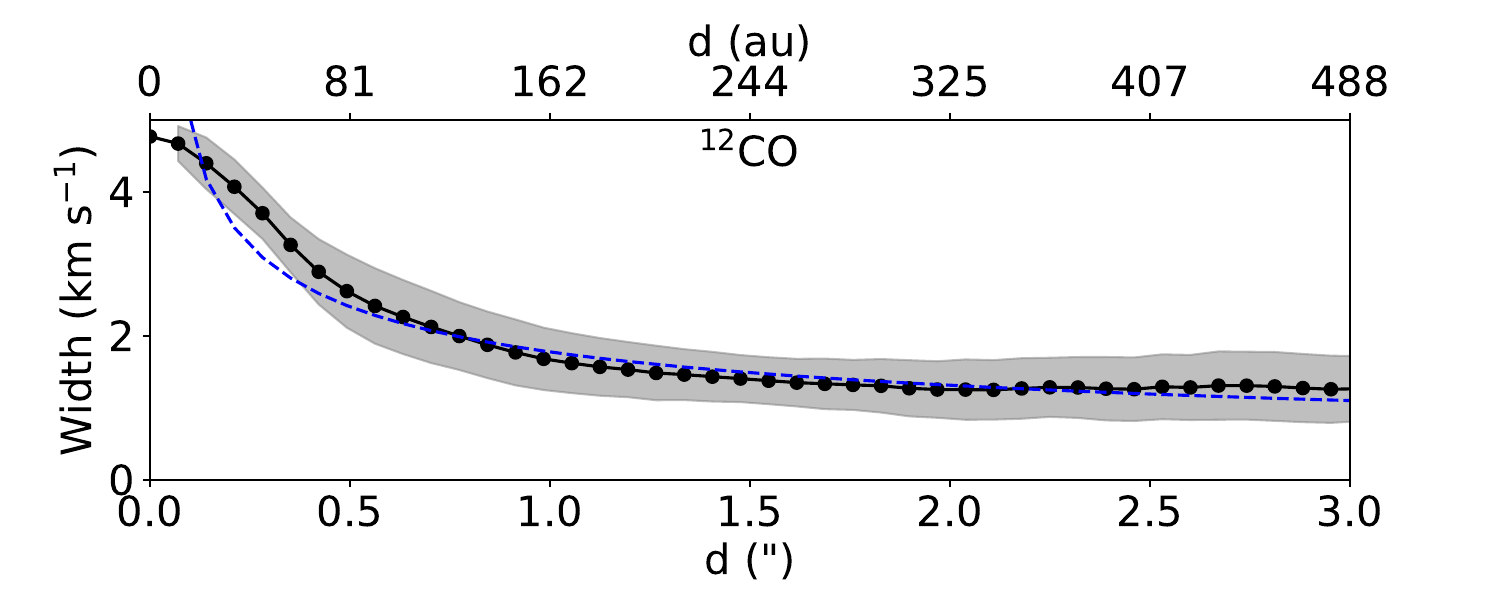}
  \includegraphics[width=0.3\textwidth, trim=0mm 0mm 0mm 0mm, clip]{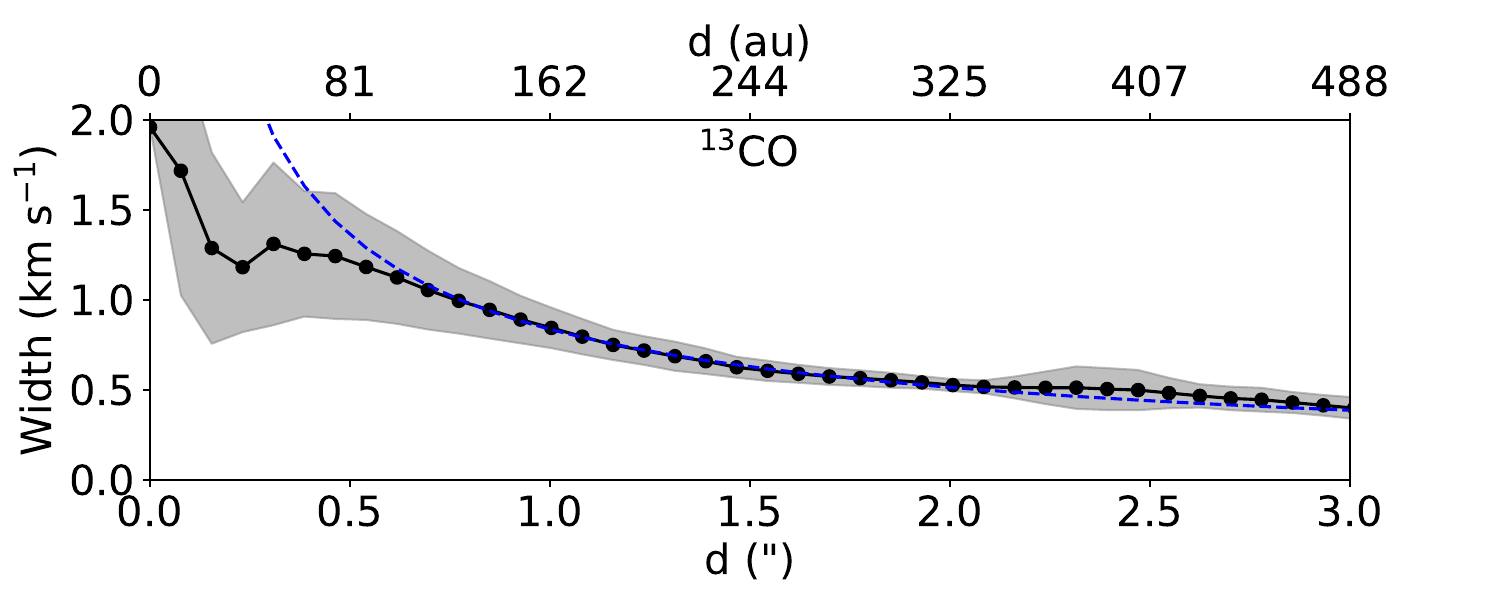}\\
  \includegraphics[width=0.3\textwidth, trim=0mm 0mm 0mm 0mm, clip]{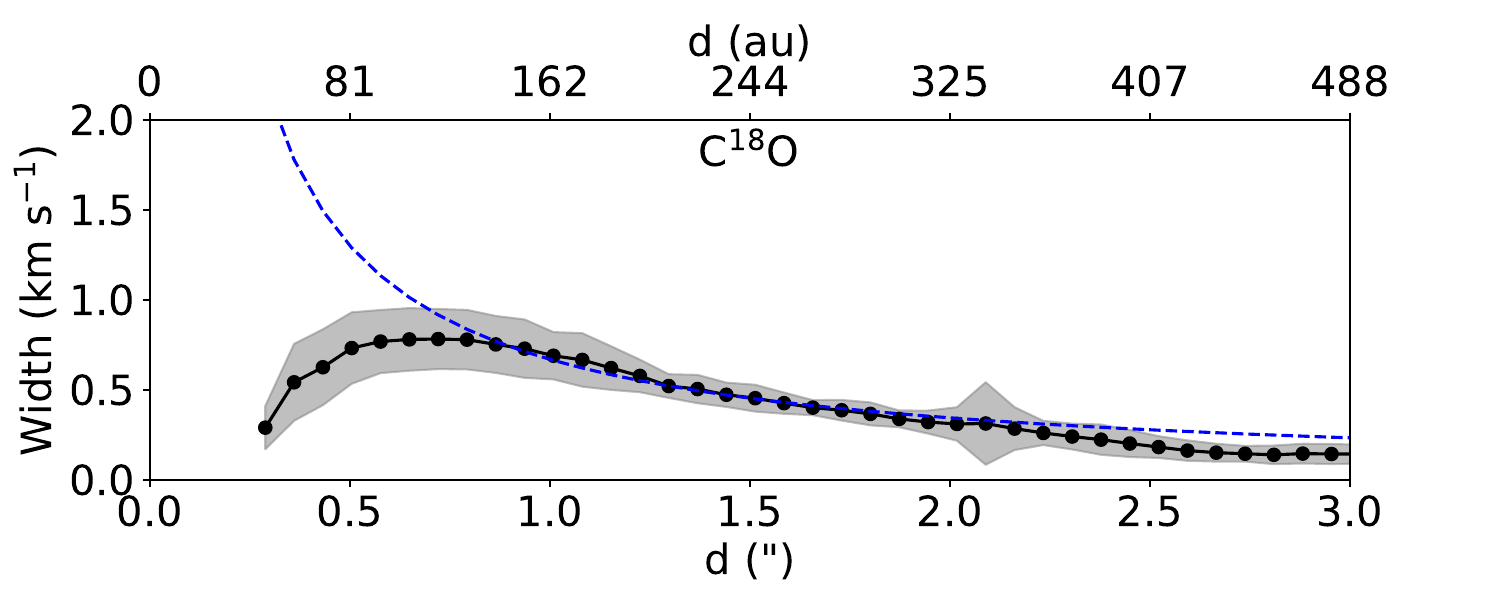}
  \includegraphics[width=0.3\textwidth, trim=0mm 0mm 0mm 0mm, clip]{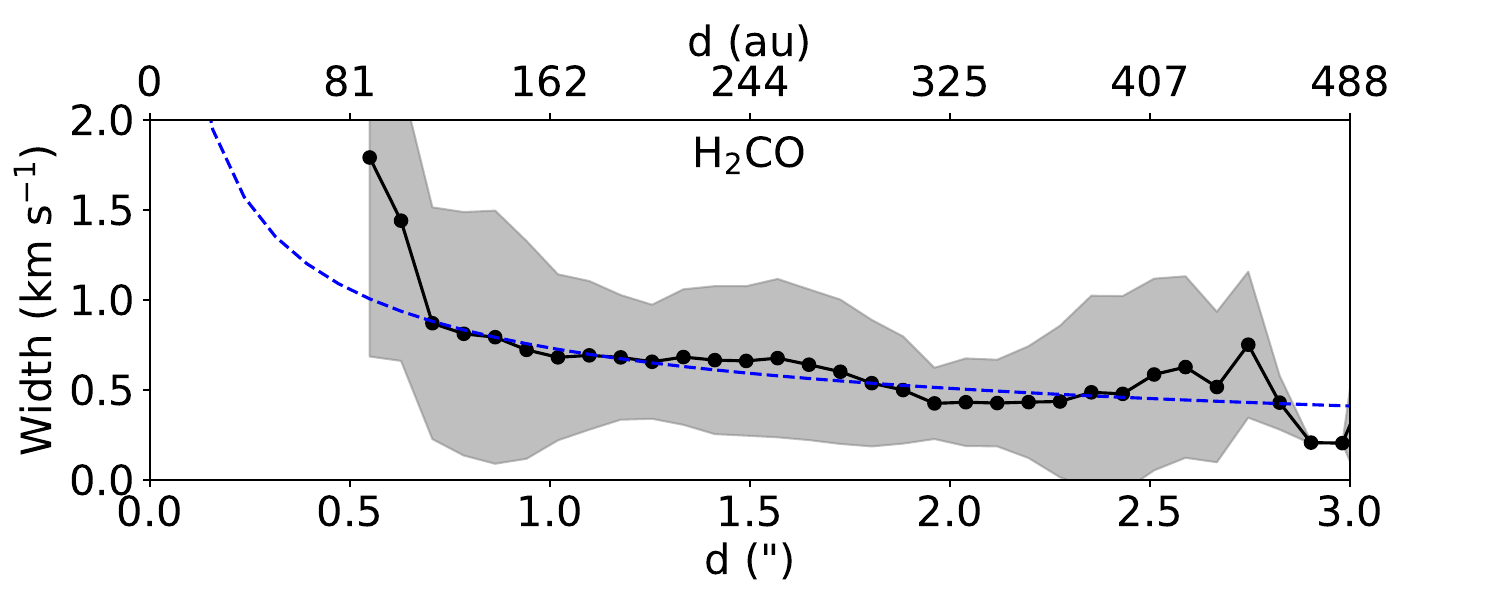}
  \includegraphics[width=0.3\textwidth, trim=0mm 0mm 0mm 0mm, clip]{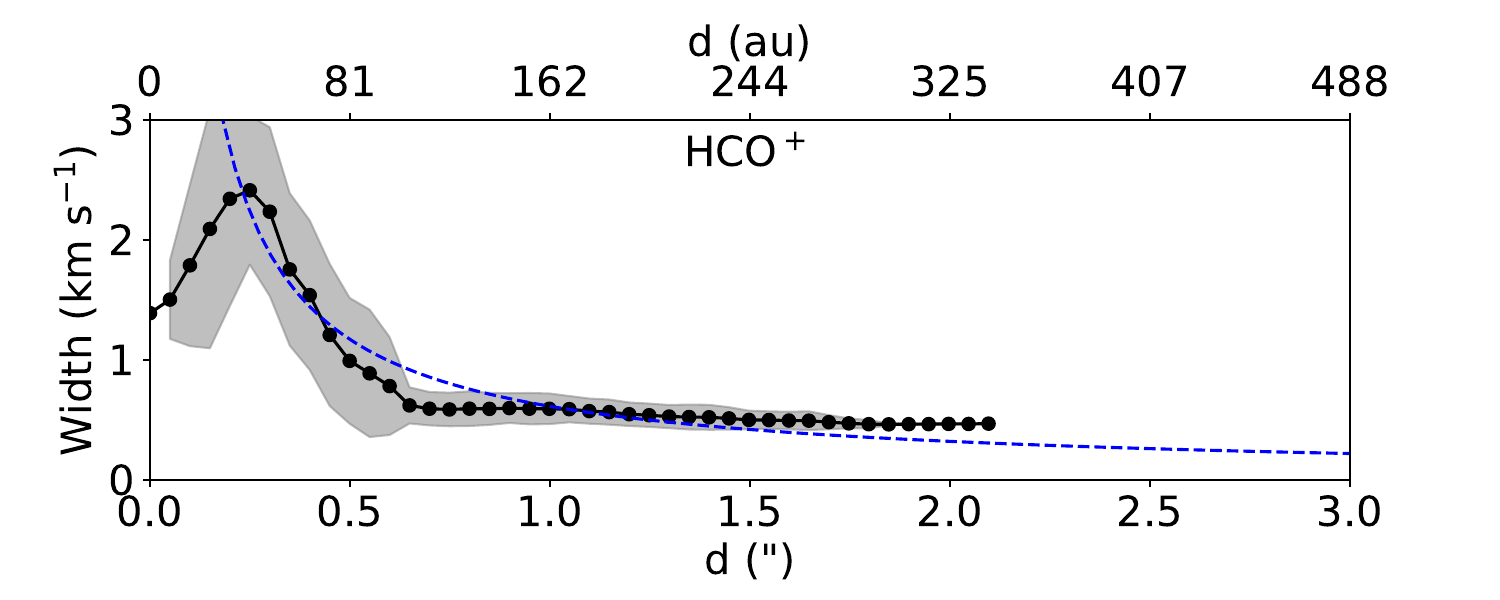}\\
  \includegraphics[width=0.3\textwidth, trim=0mm 0mm 0mm 0mm, clip]{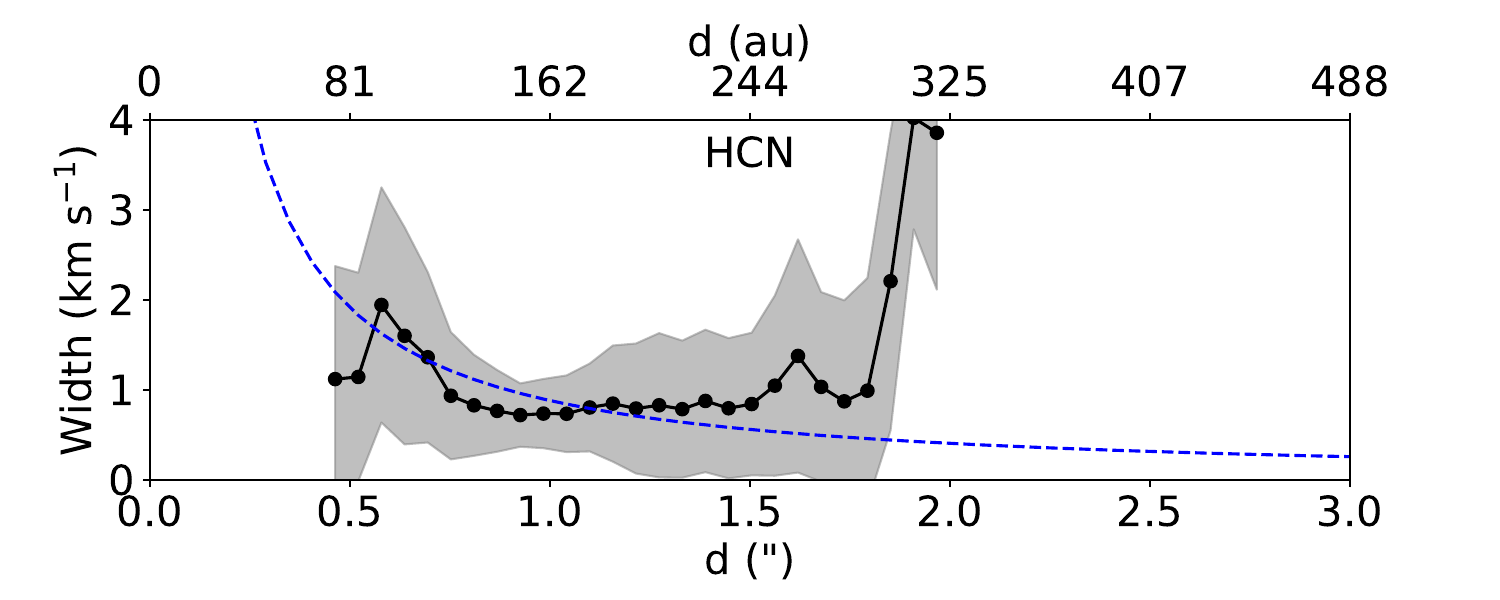}
  \includegraphics[width=0.3\textwidth, trim=0mm 0mm 0mm 0mm, clip]{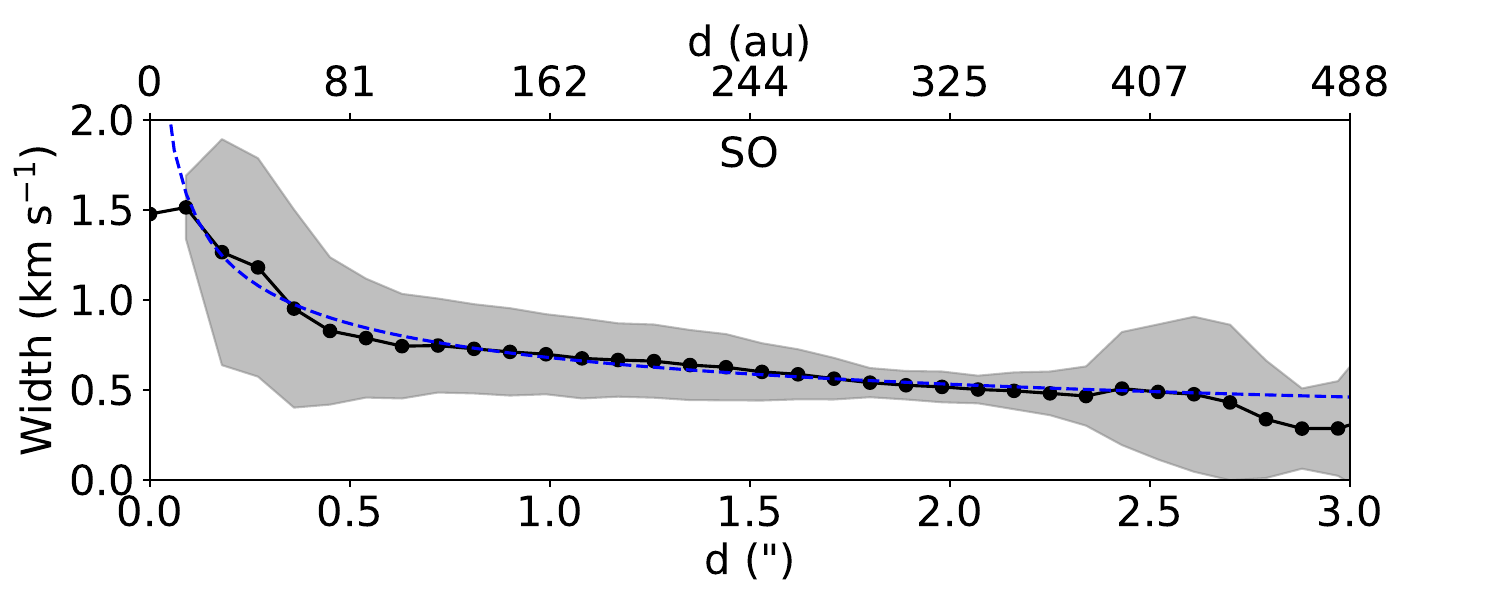}
  \includegraphics[width=0.3\textwidth, trim=0mm 0mm 0mm 0mm, clip]{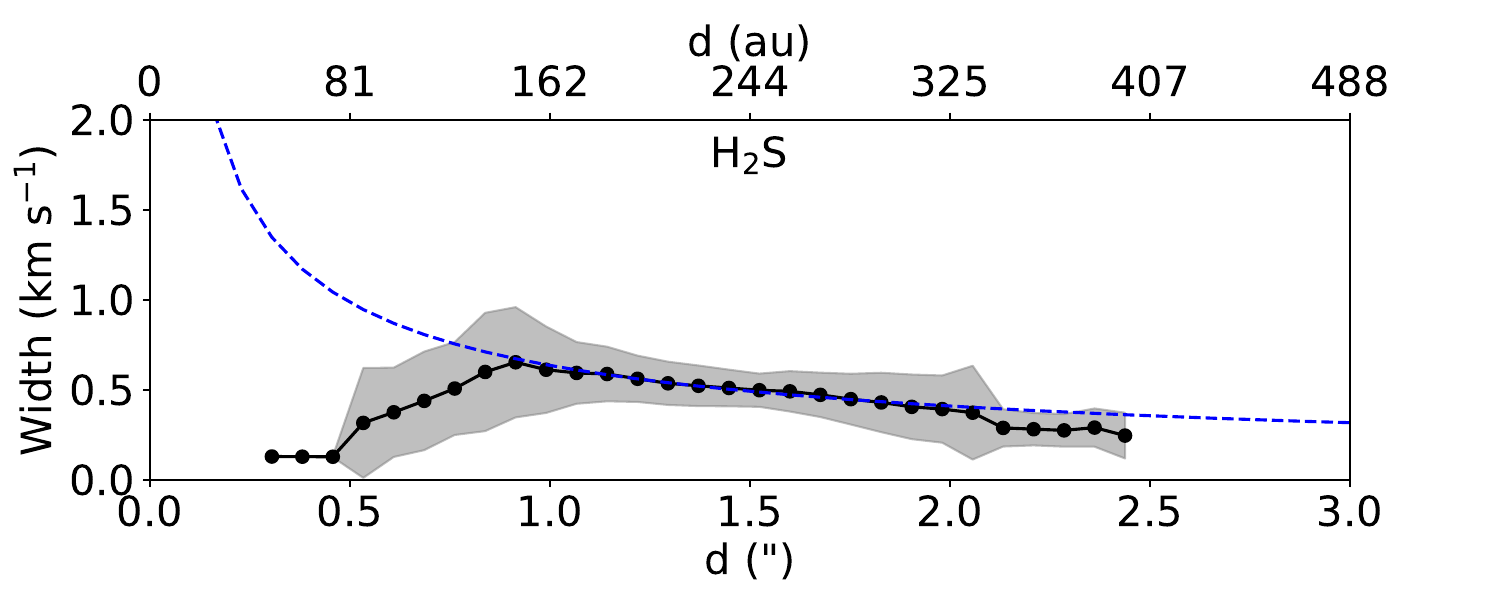}\\
  \caption{Radial profiles of the second-moment maps of the different species surveyed thus far. The blue dashed line shows a power law fit to the data. The name of each species is shown in each panel.}
 \label{Fig:width_radial_profile}
\end{center}
\end{figure*}
 
 \begin{figure}[t!]
\begin{center}
  \includegraphics[width=0.45\textwidth, trim=0mm 0mm 0mm 0mm, clip]{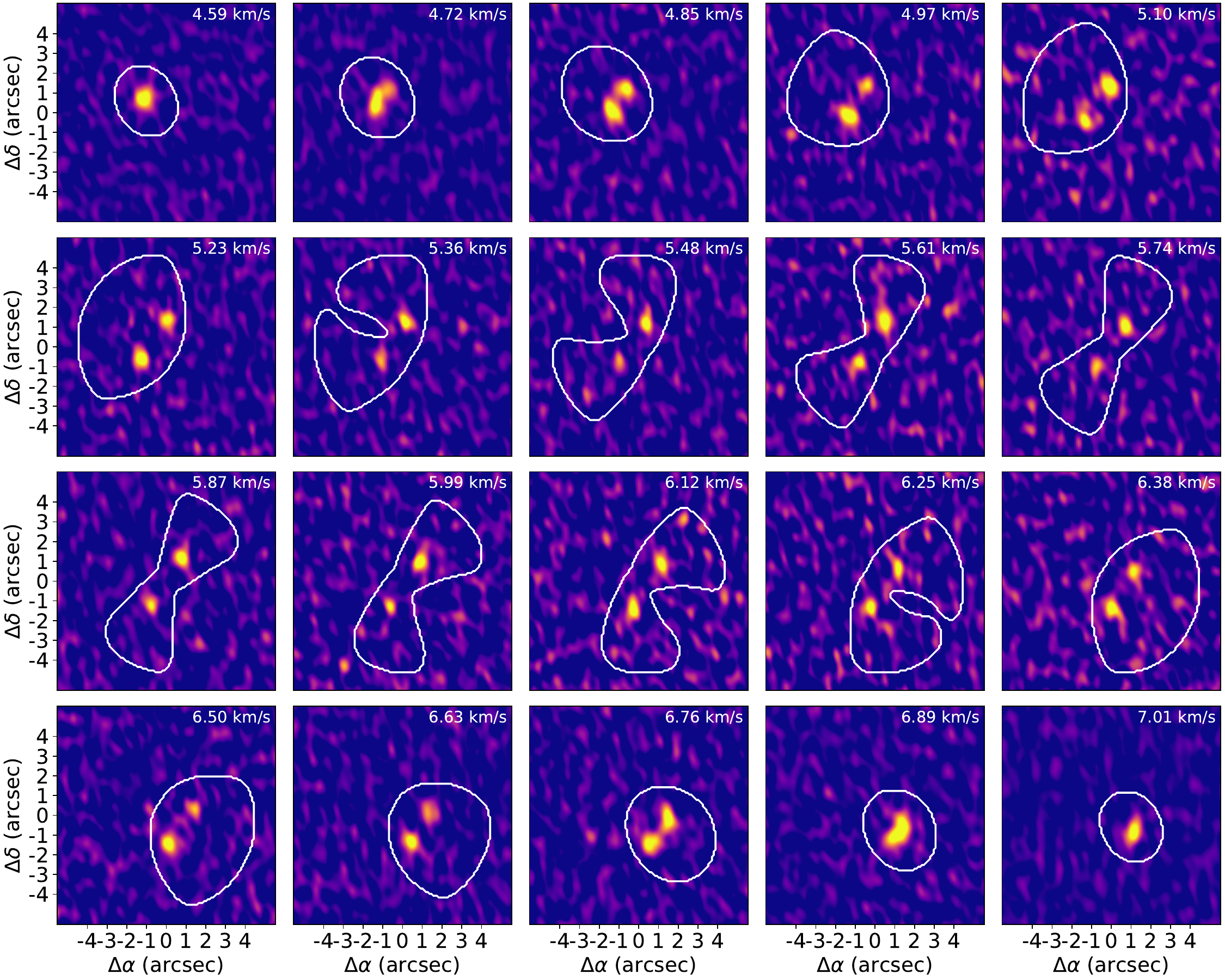}\\
  \caption{CS channel map with the Keplerian mask overlaid on each channel.}
 \label{Fig:CS_Kepler_mask}
\end{center}
\end{figure}

 \begin{figure}[t!]
\begin{center}
  \includegraphics[width=0.45\textwidth, trim=0mm 0mm 0mm 0mm, clip]{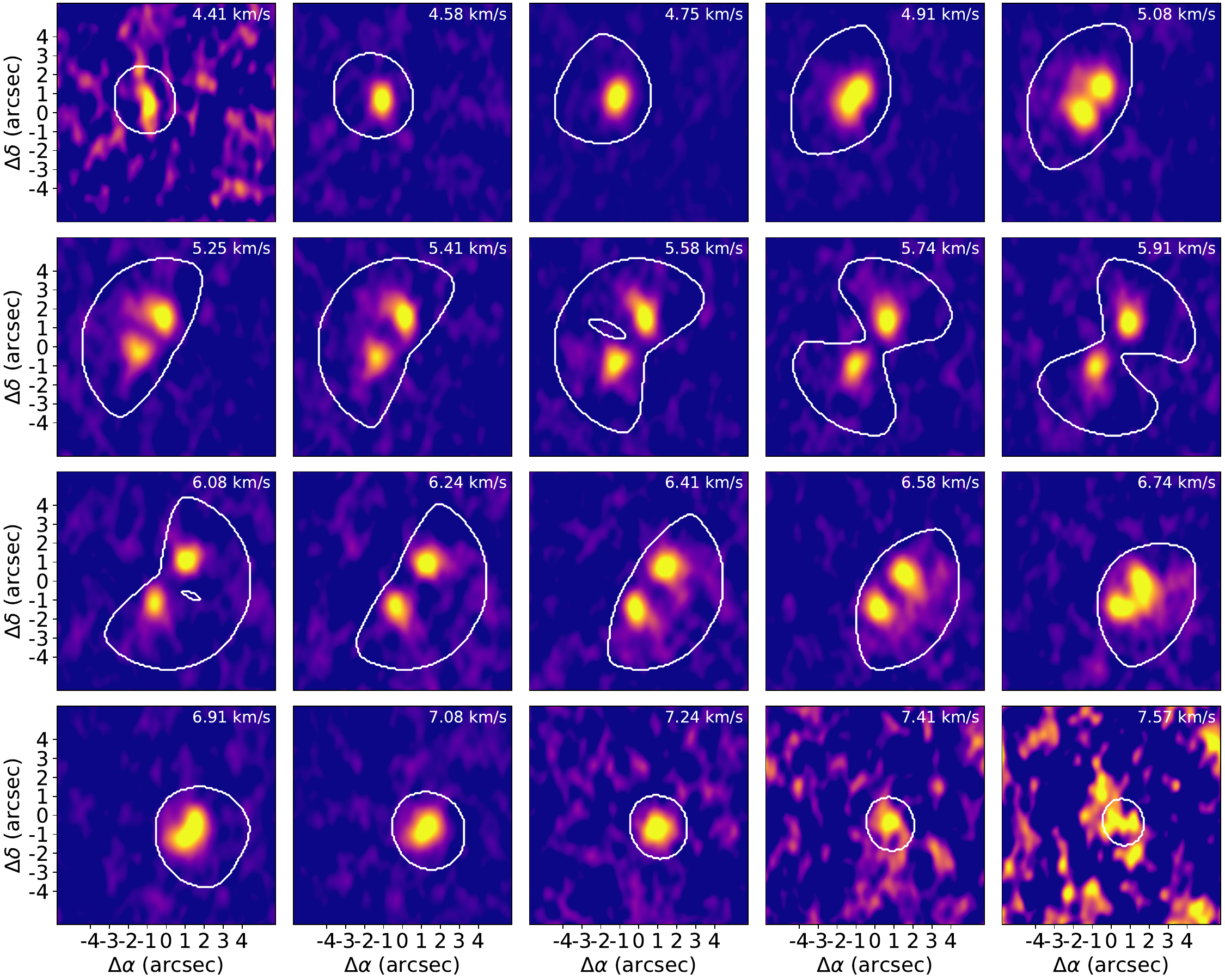}\\
  \caption{SO channel map with the Keplerian mask overlaid on each channel.}
 \label{Fig:SO_Kepler_mask}
\end{center}
\end{figure}
  
 \subsection{Keplerian masking}\label{Subsect:Kepler}
\begin{figure*}[t!]
\begin{center}
   \includegraphics[width=0.9\textwidth]{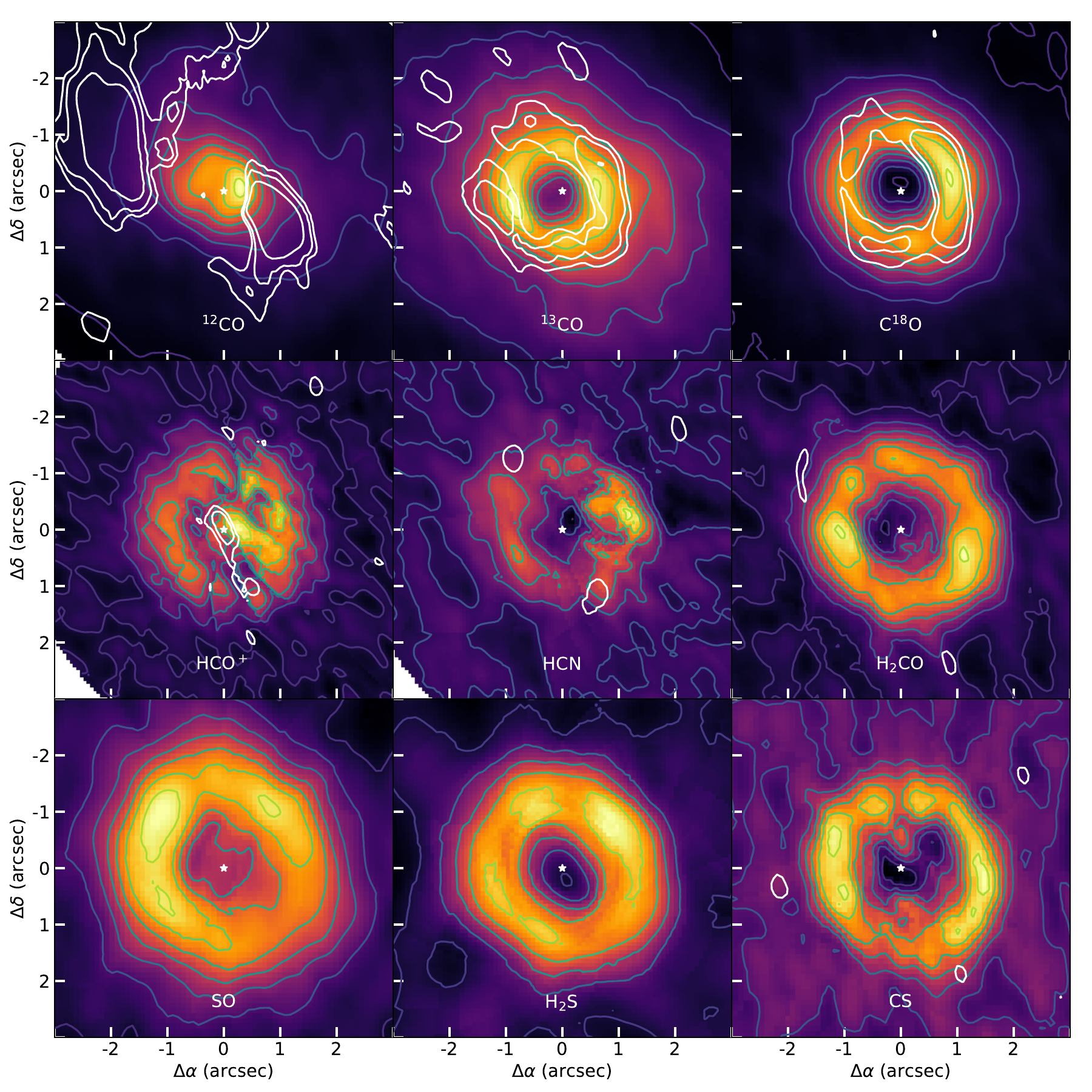}
  \caption{Integrated intensity maps of the molecular species surveyed thus far after applying a Keplerian mask (color map) with 3$\sigma$ residuals overplotted as white contours.}
 \label{Fig:Kepler_masking}
\end{center}
\end{figure*}

To help us study the kinematics of our disk, we produced a Keplerian mask using the code \texttt{Keplerian mask} presented in \cite{Teague2020}. To build the mask, the stellar mass, position angle, inclination angle, stellar distance, and system velocity must be specified. We assumed i=22.0$^{\circ}$, PA=237$^{\circ}$ from our \texttt{eddy} fits, $\rm M_{*}=$2.4M$\rm _{\odot}$, d=155.9 pc, and v$\rm _{sys}=5.8~km~s^{-1}$. \texttt{Keplerian mask} also allows us to introduce as input parameters the mean line width at 1$\arcsec$ and the power-law of the radial decay of the line width. To estimate both quantities, we produced a radial average of the line width of the deprojected map. The resulting profile is shown in Fig. \ref{Fig:width_radial_profile}, where we also show an exponential fit to the data of the form

\begin{equation}
W=W_0 \left( \frac{r}{r_0} \right) ^q
\end{equation}
where $\rm W_{0}$ is the line width at the reference radius, $\rm r_{0}=1\arcsec$, and q is the exponent of the power law. Our fit resulted in $\rm W_0 = (404\pm22) ~m~s^{-1}$, $\rm q=-0.65\pm 0.13$. We used only data points in the range 1$\arcsec$ to 2$\arcsec$ to avoid the influence of the compact source at the center. These values for the mean line width at 1$\arcsec$ and the power-law index were also fed into the Keplerian mask model. Last, the code allows the computation of the Keplerian mask at different heights above the mid-plane. In \cite{Riviere2022}, it was shown that the CS mostly originates in a region 15 au above the mid-plane when r=200 au. Thus, we assumed z/r=0.075. The resulting Keplerian mask for CS is shown in Fig.\ref{Fig:CS_Kepler_mask} overlaid on a channel map. Fig. \ref{Fig:SO_Kepler_mask} shows the same for SO for comparison. No emission is detected outside the Keplerian masks. The Keplerian masks were used to compute the integrated intensity maps shown in Fig. \ref{Fig:Kepler_masking}, where we plot the zeroth moment maps after Keplerian masking (color map), as well as the residuals from Keplerian masking (white contours). As can be seen, CS emission is Keplerian, with only three spots showing non-Keplerian residuals at a 3$\sigma$ level. We applied the same analysis to a few molecular species that our team had previously observed, including $\rm ^{12}$CO, $\rm ^{13}$CO, C$\rm ^{18}$O, H$\rm _2$CO, HCO$\rm ^+$, HCN, H$\rm _2$S, and SO. The SO 5$\rm _6$-4$\rm _5$ map analyzed in this paper is the ALMA one presented in \cite{Dutrey2024}, which had better sensitivity than the NOEMA map discussed in \cite{Riviere2020}. H$\rm _2$CO and HCN show a distribution of non-Keplerian residuals similar to that of CS, with two and three isolated spots, respectively. H$\rm _2$S and SO do not show non-Keplerian features to the 3$\sigma$ level. HCO$\rm ^+$ shows non-Keplerian residuals toward the center and along the North-South direction. C$\rm ^{18}$O, $\rm ^{13}$CO, and $\rm ^{12}$CO show prominent non-Keplerian motions along all azimuths. In the case of $\rm ^{12}$CO, the residuals trace the spiral arms detected in previous IR and mm surveys toward the source. The  C$\rm ^{18}$O and $\rm ^{13}$CO residuals show very similar distributions.

The ratios of the emission peaks in the Keplerian masked maps over the emission peak in the non-Keplerian residual maps are 6, 4, 3, 1, 0.5, and 1.0 for $\rm ^{12}CO$, $\rm ^{13}CO$,  C$\rm ^{18}$O, HCO$\rm ^+$, CS, and HCN respectively. For the species with no residuals, the ratios are 1.4 (H$\rm _2$S) and 3.3 (SO). According to these numbers, non-Keplerian residuals might still be present but undetected in H$\rm _2$S, but the lack of non-Keplerian residuals in the SO maps seems solid. 

\begin{figure}[t!]
\begin{center}
   \includegraphics[width=0.45\textwidth]{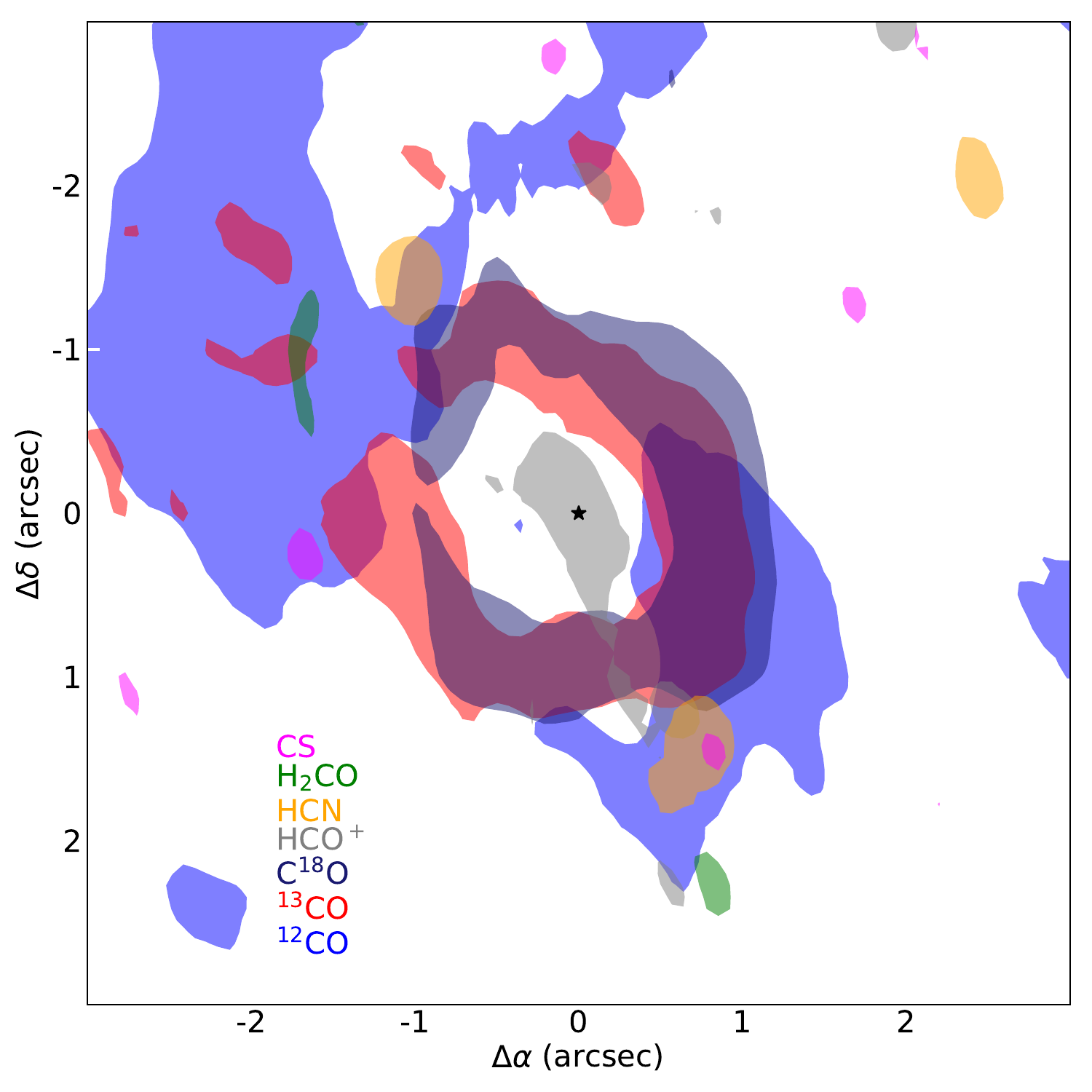}
  \caption{Distribution of non-Keplerian residuals of the different species observed.}
 \label{Fig:nonKeplerian_residuals}
\end{center}
\end{figure}

\begin{figure}[t!]
\begin{center}
   \includegraphics[width=0.45\textwidth]{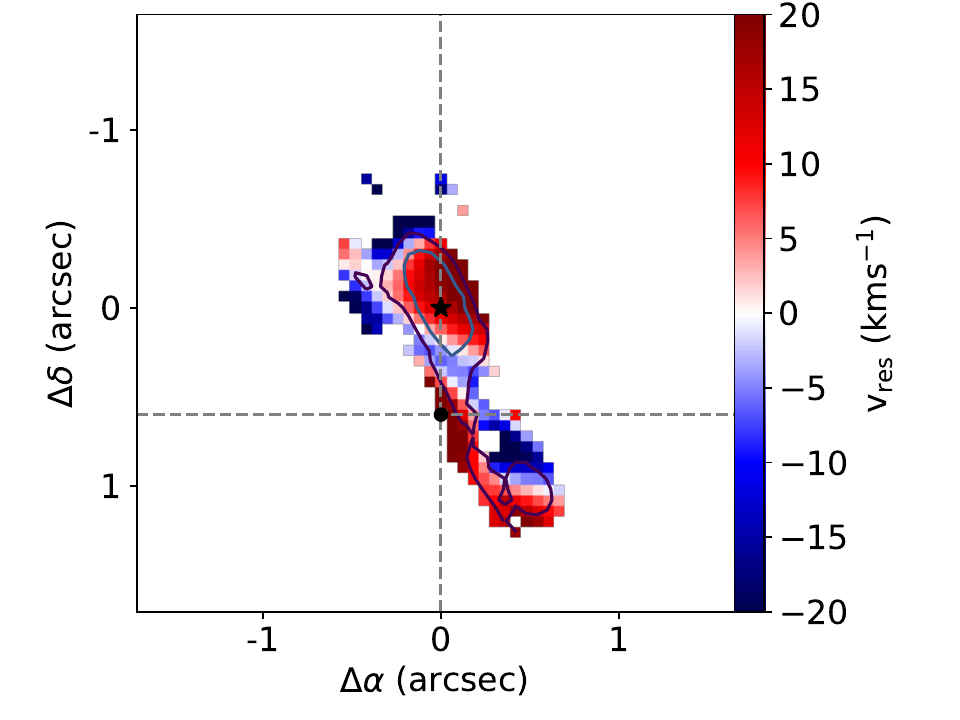}
  \caption{Velocity map of the non-Keplerian residuals in the HCO$\rm ^+$ map. The position of AB Aur b has been included for comparison.}
 \label{Fig:HCOp_planet}
\end{center}
\end{figure}

\subsection{CS column density}\label{Subsec:NCS}
Assuming LTE and optically thin emission, we can estimate the CS column density in AB Aur's disk using:

\begin{equation}\label{Eq:N}
N = \frac{N_u Q}{g_u}e^{E_u/kT_{ex}}
\end{equation}
with 
\begin{equation}
N_u = 1.94\times 10^{3} \frac{\nu^2 W}{A_{ul}}
\end{equation}
where $\rm E_u$ and $\rm g_u$ are the energy and degeneracy of the upper level, respectively, T$\rm _{ex}$ is the gas excitation  temperature, W is the velocity-integrated line intensity in units of K km s$\rm ^{-1}$, $\rm A_{ul}$ is the Einstein coefficient of the corresponding transition, and Q is the partition function:
\begin{equation}
Q = \sum_i g_i  e^{-E_i/kT}
\end{equation}
To allow the comparison with previous observations, we assumed the same gas temperature, T=30 K \citep{Riviere2022}. This temperature is consistent with the temperature derived by \cite{Pietu2005}, who estimated a kinetic temperature of $\rm \sim$30 K throughout the disk. \cite{Woitke2019} derived a mass-mean gas temperature of 54 K, and a minimum gas temperature of 26 K, close to the value we are using. 

The mean CS column density is N(CS)=(8$\rm \pm$4)$\rm \times 10^{12}~cm^{-2}$, with a maximum column density of 1.6$\rm \times 10^{13}~cm^{-2}$ and a minimum of 2.2$\rm \times 10^{12}~cm^{-2}$. We stress that the CS column density might be underestimated if the emission is optically thick. We also note that our column density estimates depend on the temperature assumed. However, the impact of such assumptions can be mitigated by observing other CS lines and using rotational diagrams to derive the gas temperature. 
  
\section{Discussion}\label{Sect:discussion}
The new CS NOEMA observations of AB Aur allowed us to resolve the CS emission at a moderate resolution. As with other species previously surveyed, CS is distributed in a ring, with emission observed roughly at distances between 0.9$\arcsec$ and 2$\arcsec$. According to our \texttt{eddy} fit, the emission is consistent with a disk with an inclination angle of 22$\rm ^{\circ}$ and a position angle of 237$\rm ^{\circ}$, matching the angle derived from observations of other species (see Table \ref{Tab:eddy_results}).

In \cite{Riviere2020}, we performed a detailed study of the spatial distribution of molecular species in the AB Aur disk, concluding that the radial emission peaks of the different species were not related to the binding energy, suggesting that thermal desorption is not the main driver of chemical segregation in the disk. The CS 3-2 ring observed with NOEMA peaks at $\rm \sim$1.3$\arcsec$, in between the SO peak at 1.4$\arcsec$ \citep{Riviere2020} and the p-H$\rm _2$CO and H$\rm _2$S peaks at 1.2$\arcsec$ \citep{Riviere2020, Riviere2022}. 
The binding energies for CS, SO, and H$\rm _2$S are 3200 K, 2700 K, and 2800 K, respectively \citep{Wakelam2017}. If thermal desorption was the main desorption mechanism, the radius of the midplane snow line would be expected to anti-correlate with the binding energy. However, no correlation was found, indicating that thermal desorption is not shaping the chemical composition and distribution in the AB Aur disk. Our new NOEMA CS observations further strengthen this idea.

\subsection{Non-Keplerian motions}
Our analysis in Sect. \ref{Subsect:Kepler} demonstrates the presence of non-Keplerian motions in the gas phase in the protoplanetary disk surrounding AB Aur (see Fig. \ref{Fig:Kepler_masking}). These non-Keplerian motions are present only in some molecular species: $\rm ^{12}$CO, $\rm ^{13}$CO, and C$\rm ^{18}$O depict strong non-Keplerian residuals at all azimuths. CS, H$\rm _2$S, and SO show a mostly Keplerian profile. In between, HCN, H$\rm _2$CO, and HCO$\rm ^+$ show localized spots with non-Keplerian motions. 

To help us with the discussion of non-Keplerian movements in the surroundings of AB Aur, we show in Fig. \ref{Fig:nonKeplerian_residuals} 3$\sigma$ filled contours of the residual emission of each species where residuals are detected. In the case of HCO$\rm ^+$, the non-Keplerian motions are observed toward the central source and along a North-South straight line. This is tracing emission along the radio jet. The HCO$\rm ^+$ non-Keplerian residuals toward the central source are likely due to a misaligned inner disk \citep{Tang2012, Riviere2019}. The HCO$\rm ^+$ residuals are also observed to the South of the central source,  close to the position of the candidate planet AB Aur b \citep{Currie2022}. This is illustrated in Fig. \ref{Fig:HCOp_planet}, where we show the velocity of the non-Keplerian residuals (color map) with overplotted integrated-intensity contours. We also show the position of AB Aur b. The residuals from HCO$\rm ^+$ emission to the South of the compact source seem to be associated with the planet. 

The non-Keplerian motions associated with $^{12}$CO, $\rm ^{13}$CO, and C$\rm ^{18}$O seem to be associated with the large-scale spirals observed in \cite{Tang2012}. Since the first detection of a large-scale streamer \citep{Pineda2020}, their presence in young stellar objects has been widely discussed \citep{Pineda2022}. The $^{12}$CO residuals are co-spatial with the spiral arms, and thus they trace gas in non-Keplerian motion along the arms. $\rm ^{13}$CO and C$\rm ^{18}$O residuals are observed in a horseshoe region that covers almost the whole protoplanetary disk. Both isotopologues show a peak in their residual distribution around the position of the dust trap. The non-Keplerian motions there could be due to increased turbulence in the surroundings of the dust trap. We notice that the maximum in SO emission is located at the position where the northeastern arm lands in the disk. H$\rm _2$CO also shows an emission peak close to the position.  We speculate that the streamer's impact on the protoplanetary disk warms up the disk, thus desorbing molecular species from the ice surface and resulting in a local enhancement of those species desorbed from the grains. The spiral arm can further enhance the abundance of other surface species through sputtering. This local enhancement will increase the overall abundance of the species desorbed. 

The analysis of non-Keplerian motions in AB Aur points to H$\rm _2$S, H$\rm _2$CO, and CS as good tracers of the Keplerian motions in a protoplanetary disk, while the rest of the species are more affected by the perturbations induced by large-scale streamers and/or spiral arms. This might be due to these species originating from deeper layers in the disk. In their study of the Flying Saucer, \cite{Dutrey2017} showed that CS emission was more confined towards the disk mid-plane than CO. This agrees with predictions by chemical models \citep{Dutrey2011}. \cite{Teague2018b} derived a low kinetic temperature consistent with an origin close to the midplane. Our results regarding Keplerian masking in AB Aur support the same idea, i.e., that CS is a better tracer of the mid-plane molecular content. Given the lack of non-Keplerian residuals in H$\rm _2$CO and H$\rm _2$S maps, the same can be said about them. 

\begin{figure*}[t!]
\begin{center}
  \includegraphics[width=0.9\textwidth, trim=0mm 0mm 0mm 0mm, clip]{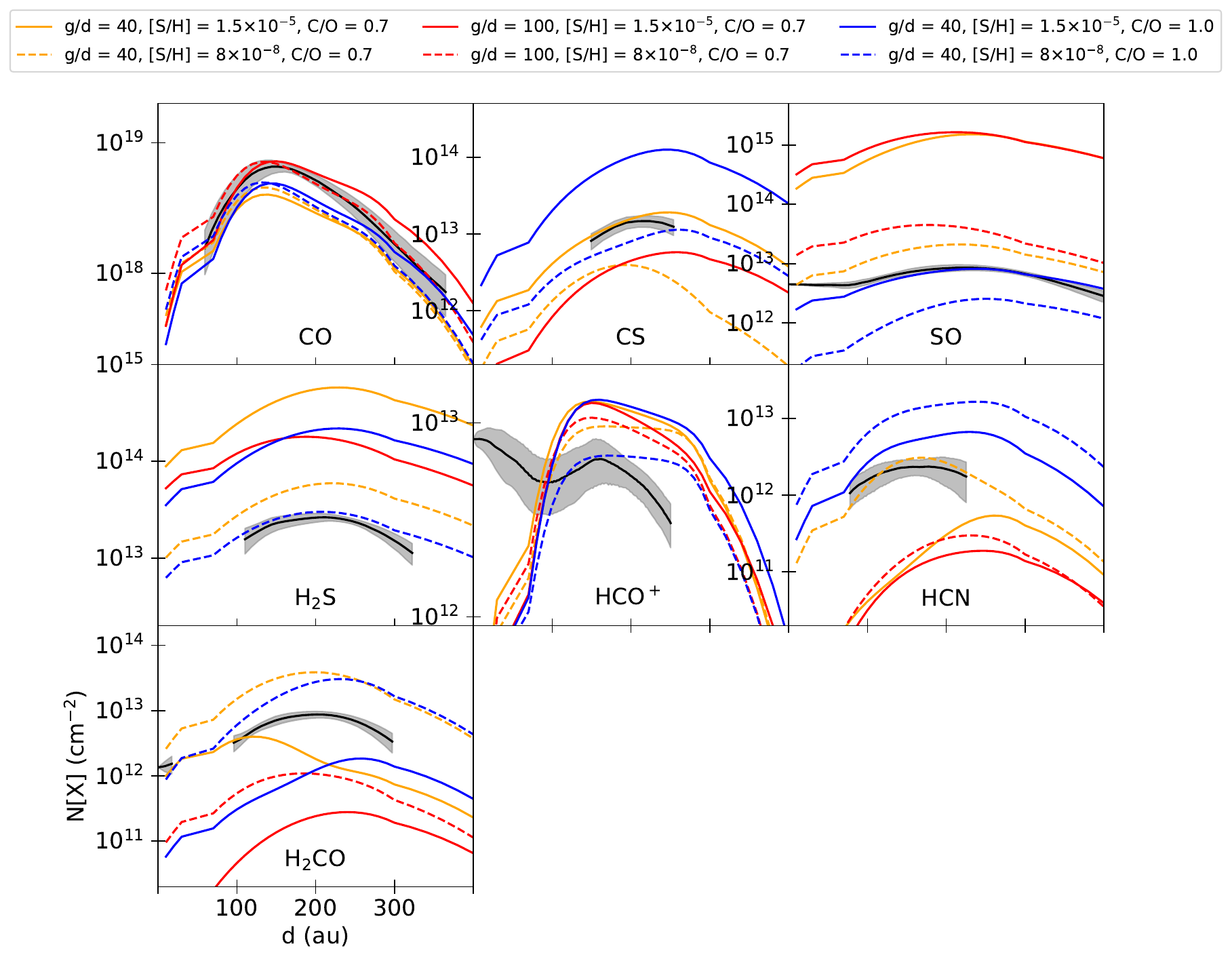}
  \caption{Comparison of observed radial profiles with NAUTILUS models from \cite{Riviere2020}. The black line is the radial profile from our observations, with the shaded area representing the standard deviation. The colored lines represent the different models (see legend).}
 \label{Fig:NAUTILUS_models_comp}
\end{center}
\end{figure*}

 \begin{table}[]
\caption{Model parameters from the models in \cite{Riviere2020}}
\label{Tab:model_setup}
\centering
\begin{tabular}{llll}
\hline \hline
Model ID & gas-to-dust mass ratio & $[S/H]$ & C/O \\
\hline
1 & 40 & 1.5$\rm \times 10^{-5}$ & 0.7 \\
2 & 40 & 8$\rm \times 10^{-8}$ & 0.7 \\
3 & 100 & 1.5$\rm \times 10^{-5}$ & 0.7 \\
4 & 100 & 8$\rm \times 10^{-8}$ &  0.7 \\
5 & 40 & 1.5$\rm \times 10^{-5}$ & 1 \\
6 & 40 & 8$\rm \times 10^{-8}$ & 1 \\
\hline
\end{tabular}
\end{table}

\subsection{NAUTILUS models}\label{Subsect:NAUTILUS}
The NOEMA observations of CS 3-2 presented in this paper can help us better understand the sulfur molecule budget in AB Aur. In \cite{Riviere2020} we compared observations of HCN, HCO$\rm ^+$, H$\rm _2$CO, SO, CO, $\rm ^{12}$CO, $\rm ^{13}$CO, and C$\rm ^{18}$O with a set of Nautilus models where the following parameters varied: the C/O ratio, the sulfur abundance, and the gas-to-dust mass ratio. The model's parameters are summarized in \ref{Tab:model_setup}. The global comparison favored models with a low gas-to-dust mass ratio, in agreement with the observational estimate of the gas-to-dust mass ratio \citep[$\rm \sim 40$,][]{Riviere2020}, which is smaller than the value of 106 derived in \cite{Woitke2019} by means of SED and line profile modelling. In their model comparison, however, the authors included observations of CO and H$\rm _2$ from the inner disk, while we only model the molecular content in the outer disk, where our observations detected molecular emission. Thus, the difference in the derived gas-to-dust ratio might arise from an enhanced ratio in the inner disk, a contribution that is not included in our disk model.  The sulfur abundance was mostly unconstrained due to the lack of observations of more sulfur-bearing species.  The comparison with HCN, H$\rm _2$CO, and SO favored models with a high C/O ratio. However, the ratio was mostly unconstrained due to the low number of molecular species included in the study. Our new CS observations will help us to better constrain the sulfur abundance and C/O abundance ratio.  In this section, we update the comparison by including observations of H$\rm _2$S \citep{Riviere2022} and CS from the present study. We also recompute the column density maps of the different species to use the same methodology used in this paper and in \cite{Riviere2022}, i. e., we estimate column densities assuming LTE. Furthermore, the column density maps are computed using the zeroth-moment maps with Keplerian masking, thus removing the contribution outside the Keplerian disk. Finally, to mimic the presence of a cavity in our models, we applied a mask such that the model values for radii smaller than $\rm R_{in}$=100 au and larger than $\rm R_{out}$=300 au, the column density is zero. The models were convolved to match the beam size of the different molecules surveyed. The resulting update is shown in Fig. \ref{Fig:NAUTILUS_models_comp}, where we compare the radial profiles of our interferometric observations with those derived from models. The inclusion of a cavity results in a better match compared to models in  \cite{Riviere2020}.

Models with a gas-to-dust mass ratio of 100 fail to reproduce most of the observed profiles and can be ruled out, the only exception being the CO profile, which is, nevertheless, also fitted by models with a gas-to-dust mass ratio of 40. Regarding the C/O ratio and sulfur abundance, the comparison is less clear, and both parameters interact in a complex way. The SO profile favors a model with a gas-to-dust mass ratio of 40, C/O=1.0, and no sulfur depletion. We note that shocks can enhance the abundance of SO locally, as SO is a well-known outflow and shock tracer \citep{Codella2003, Viti2004, Benedettini2006, Esplugues2013}. This may complicate the comparison with models since they do not include the effect of shocked gas. However, we note that the SO column density has to go down by almost one order of magnitude to match the model with C/O=1 and [S/H]=8$\rm \times 10^{-8}$. This might give us an idea of the magnitude of SO desorption from the surface of grains. H$\rm _2$S is well reproduced by models with a gas-to-dust mass ratio of 40 and sulfur depletion. CS is well matched by models with either standard sulfur abundance and C/O=0.7 or depleted sulfur abundance and C/O=1.0. We note, however, that the CS/SO ratio is more restrictive with respect to the C/O ratio. We will further discuss the model comparison regarding C/O in the following section \ref{Subsect:CS/SO_ratio}. Summarizing, there is no single model that provides a good fit to all the radial profiles, although models with a gas-to-dust mass ratio of 40 and a C/O ratio of 1 are preferred. There is no solid conclusion regarding the sulfur abundance. No model provides a satisfactory fit to the H$\rm _2$CO and HCO$\rm ^+$ profiles. Overall, the best fit is provided by the model with C/O=1, $\rm [S/H]=8\times 10^{-8}$, and a gas-to-dust ratio of 100.

The inability of astrochemical models to simultaneously reproduce H$\rm _2$S, CS, and SO has been noted previously by \cite{Navarro2020}, where the authors showed that the astrochemical model used to discuss the elemental abundances in TMC 1 and Barnard 1b failed to fit the three species mentioned simultaneously. The authors concluded that the model that better reproduces the observations was within a factor of ten of the cosmic value. This suggests that current astrochemical networks are missing important formation and destruction routes for some sulfur-bearing species. What these new routes are remains an open question. Furthermore, modeling the chemical content of a protoplanetary disk assuming a quiescent behavior might lead to wrong conclusions. In  \cite{Beitia2024} it was shown that turbulence can affect molecular abundances, especially of sulfur-bearing species. However, turbulence is typically ignored in astrochemical models, and thus, the divergence between models and observations can partially be explained by turbulence.

\subsection{CS/SO ratio}\label{Subsect:CS/SO_ratio}

\begin{figure}[t!]
\begin{center}
  \includegraphics[width=0.5\textwidth, trim=0mm 0mm 0mm 0mm, clip]{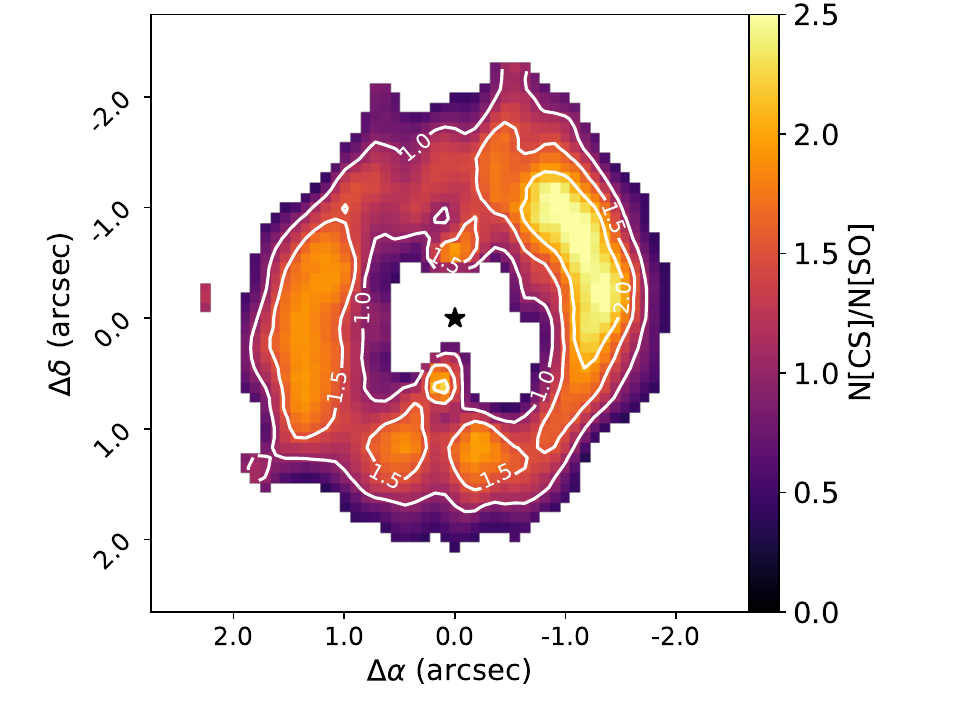}
  \caption{N(CS)/N(SO) map. Line contours mark CS/SO ratios of 1, 1.5, and 2. Only pixels where both CS and SO are detected at the 3$\sigma$ level were included.}
 \label{Fig:CS_SO_ratio_map}
\end{center}
\end{figure}

The C/O ratio is a key parameter in astrochemical models of protoplanetary disks and can be used to infer where planets were born within a protoplanetary disk \citep{Madhusudhan2012, Oberg2011}. Furthermore, together with the Mg/Si ratio, it is the most important ratio to determine the mineralogy of planets formed in it \citep{Bond2010}. The global planetary composition is set by the amount of gas versus solids accreted during the formation of the planet from different regions in the disk \citep{Madhusudhan2019}, and the composition of these solids and gas is set by the distribution of snow lines for the different volatiles (H$\rm _2$O, CO, and CO$\rm _2$, among others) in the protoplanetary disk. The link between protoplanetary disk and planetary composition is, however, bi-directional, since planet formation can also alter the chemical composition in the surroundings of the formation spot \citep{Jiang2023}. 

\cite{Bergin2016} used observations of C$\rm _2$H to constrain the C/O ratio through comparison with astrochemical models, since the abundance of C$\rm _2$H increases drastically when C/O$>$1. Alternatively, the observed CS/SO ratio can be compared with astrochemical models with different C/O ratios \citep{Dutrey2011,Semenov2018,LeGal2021,Huang2024}. \cite{LeGal2021} showed that the CS/SO ratio varies by four order of magnitude when the C/O ratio goes from 0.5 to 1.5. Following the latter approach, we studied the C/O ratio by comparing the CS observations presented in this paper with SO observations from \cite{Dutrey2024}. In Sect. \ref{Subsec:NCS} we derived the CS source-averaged column density as well as the CS column density radial profile assuming LTE and optically thin emission. To compare our CS and SO column density maps, we first computed again the SO column density map using observations from \citep{Dutrey2024} and equation  \ref{Eq:N}. We show in Fig. \ref{Fig:CS_SO_ratio_map} the resulting N(CS)/N(SO) ratio map. Most of the disk shows CS/SO$>$1.0, and the region surrounding the dust trap reaches CS/SO$>$2.0. We note that the precise value of the ratio depends on the temperature assumed (T=30 K in our case). We also note that there are large variations with azimuth, from CS/SO$>$2.0 in the dust trap to 1.0 (see Fig. \ref{Fig:CS_SO_ratio_map}). These variations could be linked to changes in the irradiation regime. \cite{Keyte2023} noted that variations in the C/O ratio in a planet-forming disk might arise from overdensity regions casting shadows on the outer parts of the disk, resulting in a decrease in the gas temperature and enhanced freeze-out of certain molecular species, such as H$\rm _2$O \citep{Esplugues2019}. The protoplanetary disk around AB Aur has long been known to be warped \citep{Hashimoto2011,Tang2012,Speedie2025}, which results in variations in the irradiation regime along the azimuth that can explain the variations in the C/O ratio.

\begin{figure}[t!]
\begin{center}
  \includegraphics[width=0.5\textwidth, trim=0mm 0mm 0mm 0mm, clip]{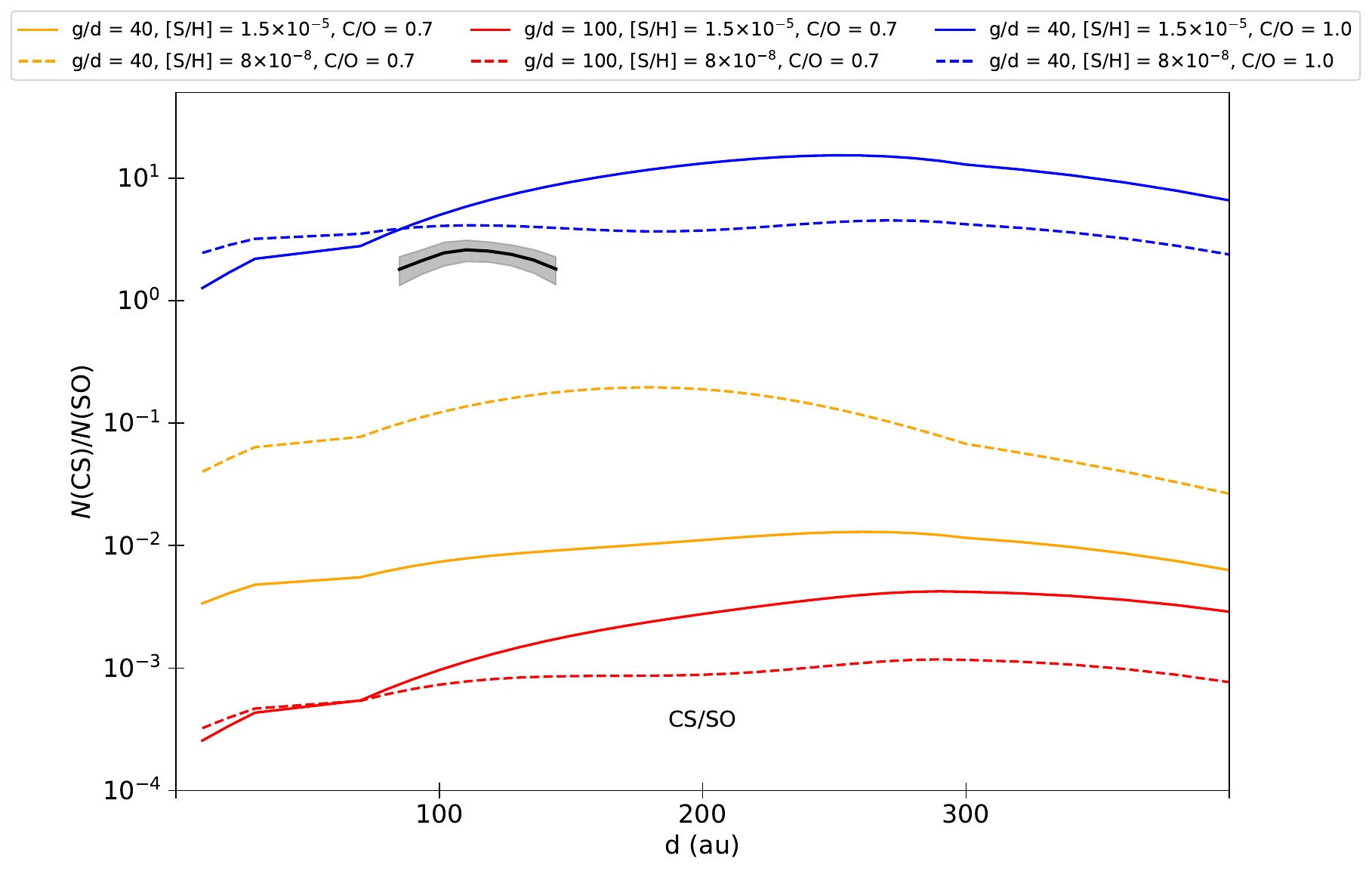}
  \caption{N(CS)/N(SO) ratios from NAUTILUS models from \cite{Riviere2020}. The black line shows the observational N(CS)/N(SO) ratio.}
 \label{Fig:NAUTILUS_CS_SO_ratio}
\end{center}
\end{figure}

A low CS/SO ratio was reported for DR Tau \cite[CS/SO$\rm \sim$0.4-0.5, ][]{Huang2024}. In contrast, \cite{Semenov2018} derived CS/SO $\rm \gtrsim$ 1 for DM Tau. \cite{LeGal2021} computed ratios in the range 4-14 or larger for five stars with spectral types A1 to K6. A CS/SO$>$1 was computed for HD 142527 \citep{Temmink2023}. Based on the non-detection of SO, \cite{Facchini2021} reported an even larger value, CS/SO$\rm >$109. Our modeling puts AB Aur in the group of disks with high CS/SO ratios, with values in the range 1.8 to 2.6.  Thus, the CS/SO ratio shows strong bimodality, with most systems showing CS/SO$\rm >$1, and a small fraction showing CS/SO$\rm <$1. Low CS/SO ratios have been observed in outflows. \cite{Bachiller1997} observed L1527 with the IRAM 30m telescope and reported column densities for a set of molecular species, including CS and SO. The CS/SO ratios at different positions in the outflow range from 0.3 to 0.9. Broadly speaking, shocks are expected to enhance SO emission \citep{Pineau1993}, thus resulting in lowered C/O ratios.

\cite{Booth2023} studied the radial distribution of the CS/SO ratio in \object{HD 169142} and noticed that the CS/SO ratio increased outwards from 1 to 12. \cite{Huang2024} reported a more complex behavior, with the ratio rising from $\rm \sim$0.3 at the center to $\rm \sim$0.6 at 100 au, then falling again to $\rm \sim$0.4 at 175 au. Our radial profile shows a flatter behavior, with the ratio increasing from 1.8 at $\rm \sim$80 au to 2.6  at 110 au and decreasing afterward to 1.8 at 144 au. 

Previous studies of the CS/SO ratio in protoplanetary disks indicate that CS/SO ratios larger than one require C/O$\rm >$1 \citep{Semenov2018,LeGal2021}. We show in Fig. \ref{Fig:NAUTILUS_CS_SO_ratio} the N(CS)/N(SO) ratio for the models from \cite{Riviere2020}. Only models with C/O$\rm =$1 result in N(CS)/N(SO) ratios compatible with our observations. Thus, the CS/SO ratio derived for AB Aur suggests C/O$\rm >$1 and sub-solar sulfur abundance. We note that including molecular ratios in the comparison provides more restrictive constraints than molecular abundances alone. 

\cite{Oberg2011} proposed a simple disk model to study the radial dependence of the C/O ratio, and derived a C/O ratio of 1 for distances larger than 10 au. Broadly speaking, the C/O ratio is controlled by the relative distance of the forming planet with respect to the central star and the water snow line at the time when the planet is assembling its atmosphere \citep{Cridland2016}. We have derived a CS/SO ratio that increases from 1.4 at the innermost parts of the disk to 1.8 at $\rm \sim$200 au, to go down afterwards to 1.3 at 250 au, in agreement with theoretical predictions \citep{Oberg2011}. Thus, a planet formed in the remaining AB Aur disk will, in principle, have C/O$>$1, i.e., will have an enriched C content and thus will have a different composition and mineralogy from Solar System planets \citep{Bond2010,Madhusudhan2011}. We note, however, that planets are born in the disk midplane, while the C/O ratios that we derive mostly trace the warm molecular layer. Furthermore, the composition of the midplane is barely constrained by observations \citep{Eistrup2018}, since the emission of most molecular species originates in the warm molecular layer. Planets with C/O ratios higher than the solar value have been observed before \citep{Madhusudhan2011,Teske2013}. Thus, these carbonaceous planets seem to be common and result from accreting their atmospheres in a region where C/O$>$0.8 \citep{Bond2010}. These planets can even form around host stars with C/O$<$1, since observations have shown that the C/O ratio in planetary atmospheres can be different from that of the host star \citep{Madhusudhan2012, Teske2013}. The planet candidate AB Aur b \citep{Currie2022} is located at $\rm \sim$ 93 au from, and thus its composition is also likely dominated by carbonaceous materials, as proposed for HD 169142 b \citep{Booth2023} based on its location. However, planets can migrate during their formation and evolution, and thus the radial distance of AB Aur might have changed since the moment it was born.

\section{Summary and conclusions}\label{Sect:summary}
We presented new NOEMA observations of CS 3-2 emission toward AB Aur. The emission originates in a ring extending from $\rm \sim$0.9$\arcsec$ ($\rm \sim$140 au) to $\rm \sim$2.0$\arcsec$ ($\rm \sim$311 au), with an emission peak at 1.3$\arcsec$ ($\rm \sim$203 au), and an azimuthal contrast ratio of 2.4. In contrast, the 2mm continuum disk extends from 0.6$\arcsec$ ($\rm \sim$93 au) to 1.5$\arcsec$ ($\rm \sim$234 au), and shows a similar azimuthal contrast ratio of 2.3. The integrated intensity emission map of CS shows two local maxima, one close to the dust trap,  and one at roughly 180$\rm ^\circ$ in azimuth from the dust trap. In the following, we summarize our main results:

\begin{itemize}

\item We have used \texttt{eddy} to fit the first-moment map of the different molecular species we have observed thus far with NOEMA aiming to derive an accurate value of the inclination and position angles. The mean values are i=22.0$\rm ^{\circ} \pm$0.5$\rm ^{\circ}$ and PA=237.0$\rm ^{\circ} \pm$0.7$\rm ^{\circ}$.

\item We find no correlation between the radius of the mid-plane snowline and the binding energy of the molecules surveyed thus far, which indicates that thermal desorption is not shaping the chemical composition and distribution in the AB Aur disk.

\item We constructed a Keplerian mask to study non-Keplerian motions in the CS 3-2 integrated intensity map. No residuals were found at the 3$\sigma$ level. The presence of non-Keplerian residuals in the integrated intensity maps of other species might be due to the impact of the streamer desorbing molecular species from the surface grains, increasing their abundance in the gas phase. The absence of non-Keplerian residuals in CS suggests that this emission originates deeper in the disk. In line with this result, we propose that CS is a better tracer of the quiescent Keplerian disk than C$\rm ^{18}$O.

\item We derive a CS/SO ratio of 1.8 to 2.6 and find that only models with C/O$\sim$1 reproduce these values. We note, however, that no model can simultaneously reproduce the CS/SO ratio and the molecular column densities. The lack of agreement between models and observations is likely due to missing formation and destruction routes for sulfur-bearing species.

\item The comparison with Nautilus models favors a model with C/O=1.0, a gas-to-dust mass ratio of 40, and [S/H]=8$\rm \times 10^{-8}$.

\item The C/O ratio derived in this study means that planets formed in the remaining AB Aur disk will have elevated atmospheric C/O ratios, and that the planet candidate AB Aur b is also most likely a carbon-rich planet, different from the planets in the Solar System.
\end{itemize}

\begin{acknowledgements}
A.F., G.E., and P.R.M. are members of project PID2022-137980NB-I00, funded by MCIN/AEI/10.13039/501100011033/FEDER UE.  This project has received funding from the European Research Council (ERC) under the European Union's Horizon Europe research and innovation program ERC-AdG-2022 (SUL4LIFE , GA No 101096293). D.S. has received funding from the European Research Council (ERC) under the European Union's Horizon 2020 research and innovation programme (PROTOPLANETS, GA No. 101002188). A.S.M. acknowledges support from ANID / Fondo 2022 ALMA / 31220025. Part of this work was supported by the Max-Planck Society. This paper makes use of the following ALMA data: ADS/JAO.ALMA\#2019.1.00579.S and ADS/JAO.ALMA\#2021.1.00690.S. This work is based on observations carried out under project number W21BA with the IRAM NOEMA Interferometer. IRAM is supported by INSU/CNRS (France), MPG (Germany) and IGN (Spain).
\end{acknowledgements}

 \bibliographystyle{aa} 
\bibliography{biblio}

\end{document}